\pdfoutput=1
\documentclass[%
 pre,amsmath,amssymb,superscriptaddress,
 reprint,%
]{revtex4-1}
\usepackage{hyperref}
\usepackage{amsfonts}
\usepackage{txfonts}
\usepackage{graphicx}
\usepackage{bm}
\usepackage{ctable}
\usepackage{url}
\usepackage{breakurl}
\usepackage{mathptmx}
\usepackage{etoolbox}
\usepackage{fixltx2e} 
\usepackage[]{cleveref}
\crefname{section}{\S\!}{\S\S\!}
\Crefname{section}{Section}{Sections}
\crefname{appendix}{App.}{Apps.}
\Crefname{appendix}{Appendix}{Appendices}
\crefname{equation}{Eq.}{Eqs.}
\Crefname{equation}{Equation}{Equations}
\crefname{figure}{Fig.}{Figs.}
\Crefname{figure}{Figure}{Figures}

\newcommand{\B}{\bm{B}}
\newcommand{\ba}{\bm{b}} 

\newcommand{\mB}{\overline{\bm{B}}}
\newcommand{\va}{\bm{v}_{\rm A}}
\newcommand{\vah}{\hat{\bm{v}}_{\rm A}}
\newcommand{\vam}{{v}_{\rm A}}
\newcommand{\tps}{\Phi} 
\newcommand{\p}{\bm{p}}
\newcommand{\ph}{\hat{\bm{p}}}
\newcommand{\tpsi}{\Phi_0} 
\newcommand{\tpb}{\vartheta} 
\newcommand{\tp}{\theta_p} 
\newcommand{\tpi}{\theta_{p0}} 
\newcommand{\tppi}{\theta_{\perp0}} 
\newcommand{\mA}{\overline{A}}
\newcommand{\zades}{J+22} 
\newcommand{\alfreds}{M+21}

\makeatletter
\def\@email#1#2{%
 \endgroup
 \patchcmd{\titleblock@produce}
  {\frontmatter@RRAPformat}
  {\frontmatter@RRAPformat{\produce@RRAP{*#1\href{mailto:#2}{#2}}}\frontmatter@RRAPformat}
  {}{}
}%
\makeatother
\nonstopmode
\begin{document}

\title[]{On the properties of Alfv\'enic switchbacks in the expanding solar wind: the influence of the Parker spiral }

\author{Jonathan Squire}%
 \email{jonathan.squire@otago.ac.nz}
 \affiliation{Physics Department, University of Otago, Dunedin 9010, New Zealand}
 \author{Zade Johnston}
\affiliation{Physics Department, University of Otago, Dunedin 9010, New Zealand}
\author{Alfred Mallet}
\affiliation{Space Sciences Laboratory, University of California, Berkeley, CA 94720, USA}
 \author{Romain Meyrand}
 \affiliation{Physics Department, University of Otago, Dunedin 9010, New Zealand}

\date{\today}

\begin{abstract}
Switchbacks --- rapid, large deflections of the solar wind's  magnetic field  --- have generated significant interest as possible
signatures  of the key mechanisms that heat the corona and accelerate the solar wind. In this context, an important task 
for theories of switchback formation and evolution is to understand their observable distinguishing features, 
allowing them to be assessed in detail using spacecraft data. Here, we work towards this goal
by studying the influence of the Parker spiral on the evolution of Alfv\'enic switchbacks in an expanding plasma. 
 Using simple analytic arguments based on the physics of  one-dimensional spherically polarized (constant-field-magnitude) Alfv\'en waves, 
 we find  that, by controlling the wave's obliquity, a Parker spiral   strongly impacts switchback properties.  
 Surprisingly, the Parker spiral 
 can significantly enhance switchback formation, despite normalized wave amplitudes growing more slowly in its presence.
In addition, switchbacks become strongly asymmetric:  large switchbacks preferentially involve magnetic-field rotations
 in the plane of the Parker spiral (tangential deflections) rather than perpendicular (normal) rotations, and such deflections are strongly ``tangentially skewed,''
meaning switchbacks always involve field rotations in the same direction (towards the positive-radial direction for an outwards mean field). 
 In a companion paper, we show that these properties also occur in turbulent  3-D fields with switchbacks, with various caveats.
These results demonstrate that substantial care is needed in assuming that specific features of switchbacks can be 
used to infer properties of the low corona; asymmetries and nontrivial correlations can develop as switchbacks propagate due to
the interplay between expansion and  spherically polarized, divergence-free magnetic  fields. 
\end{abstract}
\maketitle


\section{Introduction}

Early observations of switchbacks by Parker Solar Probe (PSP) provided a stark demonstration 
of the dynamic nature of the near-Sun solar wind \cite{Bale2019}. 
These strong and sudden reversals of the background magnetic field, which
are known from electron measurements to have the topology of local folds in field lines (as opposed to global polarity reversals \cite{Kasper2019}), 
 contain significant energy content compared to the background plasma. This suggests they play a 
 role in --- or are least a helpful diagnostic of --- the key processes that heat and 
 accelerate the solar wind \cite{Bale2021}. As such, a range of different explanations have been 
 put forth for their origin, each relating in some way to fundamental low-coronal or solar-wind physics with 
 the switchbacks emerging as an observable consequence. These explanations include models relating 
 to interchange reconnection near the solar surface \cite{Fisk2020,Drake2021,Zank2020,He2021}, low-coronal jets or other motions \cite{Sterling2020,Magyar2021}, 
field-line folding due to asymmetries from interchange reconnection \cite{Schwadron2021}, instabilities between different 
 wind streams \cite{Ruffolo2020}, or the growth of Alfv\'enic fluctuations due to plasma expansion \cite{Squire2020,Shoda2021,Mallet2021}.
Given these widely differing ideas, each of which provide different
predictions for switchback properties and occurrence rate,  a clear path forward to bettering our  understanding of the solar wind
presents itself: quantify various (hopefully) unique and observationally testable predictions for each model, then 
compare these to observations.

In this paper, we continue this process of better understanding model predictions for the scenario where switchbacks
form due to the growth of  Alfv\'enic fluctuations. 
This scenario, which is most naturally  associated with wave-driven solar-wind acceleration theories \cite{Cranmer2007,Chandran2021}, is in some sense the simplest of all those mentioned above --- it requires Alfv\'en waves of some 
form to be released into the low corona \cite{DePontieu2007}; in growing to normalized amplitudes ${\gtrsim}1$ due to expansion they create switchbacks. Importantly, the scenario does not assume anything in particular
about the process that creates the Alfv\'enic fluctuations in the first place, but  concerns only the 
role of expansion and the assertion that (because they are Alfv\'enic) they should be ``spherically polarized,'' \emph{viz.,} have a constant magnetic field strength (as is observed \cite{Laker2022}). Thus, 
its predictions relate exclusively to how fluctuations grow and change shape as they propagate. Accordingly, if a particular correlation, structure, or feature arises
as a basic prediction of the scenario, this would arise naturally from almost any mechanism 
that generates Alfv\'enic fluctuations at low altitudes. 
While this certainly does not eliminate the possibility that the original fluctuations result
from an interesting  mechanism  (e.g., interchange reconnection), it does undermine the credibility
of using switchbacks as evidence for the importance of this mechanism. 
Several  examples of this mindset 
have already been discussed in theory of Ref.~\onlinecite{Mallet2021} (hereafter \alfreds), which argues, for example, that
Alfv\'enic switchbacks are preferentially elongated along the background magnetic field as a simple consequence of being divergence free,
as well as predicting various non-trivial compressive correlations that result from expansion.

In this paper, we consider how  the Parker spiral --- the azimuthal rotation of the mean field with increasing heliocentric radius \cite{Parker1958} --- influences the evolution of Alfv\'enic switchbacks
in an expanding plasma. Our core results are  based on the idealized physics governing the evolution of large-amplitude, one-dimensional waves. 
We predict a number of surprising features and asymmetries that arise because of the Parker spiral's effect on the wave's obliquity. 
Despite their idealized nature, however, these results seem to be well confirmed by 
our companion paper Ref.~\onlinecite{Johnston2022} (hereafter \zades), 
which is designed to  be considered in conjunction with this work. \zades\ shares the same overarching goal 
of honing the predictions of the Alfv\'enic switchback model, but uses realistic 3-D expanding magnetohydrodynamic (MHD) 
simulations that capture the full complexity of 3-D structures, turbulence, and compressibility. Thus, despite the obvious limitations of treating 1-D waves (some of which are quite severe), the approach seems to yield useful 
dividends in straightforwardly understanding a number of seemingly perplexing properties of switchbacks in 3-D simulations.
Our hope --- backed by some preliminary evidence --- is that such conclusions apply also to the real solar wind.

The main result of this work (likewise, a key result of \zades), is that the Parker spiral 
causes switchbacks to become proncouncedly asymmetric in a number of ways. Defining the radial (R), tangential (T), normal (N) directions in the 
usual way --- the Parker spiral lies in the radial-tangential (R-T) plane with a component in the $+$T direction, while normal is perpendicular 
to the R-T plane (see \cref{fig: geometry}) --- we show, among other results, that: (i) switchbacks preferentially involve  field deflections in the R-T plane, as opposed to the normal direction (ii) switchbacks prevalence and intensity
is enhanced by the Parker spiral, and (iii)  field deflections become strongly ``tangentially skewed,'' meaning they always deflect towards the positive radial direction in an outwards mean field, as opposed to the opposite direction (more generally, they deflect towards the mean field's radial component). A corollary of point (iii) is that the most common magnetic-field direction (its mode) 
can be very different to its mean, because its probability density function is highly skewed. 
Per our discussion above, this implies that any fluctuation directional asymmetries in observed switchbacks do not necessarily signify that 
the original source of the switchbacks is asymmetric; rather, asymmetries (of our predicted form) arise organically due to plasma expansion as fluctuations propagate outwards.

The remainder of the paper is organized as follows. \Cref{sec: background} provides the necessary background to our calculations, introducing
key concepts such as the definition of a nonlinear Alfv\'en wave, and how the divergence-free and constant-magnetic-field-strength constraints determine 
its associated fluctuations parallel to the mean field. This section effectively summarizes some key results of \alfreds, which forms the basis for our work. 
\Cref{sec: SB formation due to expansion} presents results related to how the size of the  parallel field perturbation, which is loosely interpreted as the switchback prevalence, evolves in 1-D waves as they grow and decay due to expansion. We will see that the
Parker spiral has a  strong influence because of its effect on the
evolution of the wave's obliquity. \Cref{sec: structure} then considers the magnetic-field structure of switchbacks that 
evolve as discussed in \cref{sec: SB formation due to expansion}, illustrating their tangential skewness and providing
a simple proof for why this occurs in 1-D. \Cref{sec: discussion} is then concerned with the thorny question of how our 1-D results
can apply to fully 3-D fields, including the effects of 3-D structure, turbulence and parametric decay, 
with reference to many of the results from \zades. We see that the results cannot 
be universally applicable but can likely be reasonably  applied to certain regimes, so long as various important caveats are kept in mind.
We conclude in \cref{sec: conclusions}. 

Three appendices present tangential results. \Cref{sub: integral estimate} provides a different, 3-D, argument for
the main results of \cref{sec: SB formation due to expansion} based on integrating over a spectrum of waves that 
has been influenced by plasma expansion. \Cref{app: comparison to MHD} provides a cursory comparison of 
idealized wave results based on \alfreds\ to two-dimensional MHD simulations with expansion, showing mostly good agreement (and
allowing better understanding of a case where the method of \alfreds\ fails). \Cref{app: zpx sclaing}
demonstrates that, independently of the mean-field direction, wave amplitudes scale with expansion in the same way.

\section{Nonlinear Alfv\'en waves and the influence of expansion}\label{sec: background}

In this section, we discuss and derive some key properties of large-amplitude 
Alfv\'en waves, which will form the basis for our analysis of how such waves evolve in 
an expanding plasma with a Parker spiral. 
Our focus is on how such waves create switchbacks, \emph{viz.,} under what conditions they can reverse the mean field.  
Our notation used below is as follows: $P$,  $\rho$, $\bm{u}$ and $\bm{B}$ are the plasma's thermal pressure, density, 
flow velocity, and magnetic field. Where appropriate, we will denote spatial averaging 
with a bar (e.g., $\overline{\B}$) and fluctuating quantities (i.e., the remainder) with a $\delta$ (e.g., $\delta \B$),
using periodic boundary conditions.
In order to clarify notation, we will almost exclusively reference the mean field through the Alfv\'en speed, $\va \equiv \overline{\B}/\sqrt{4\pi\overline{\rho}}$, and reference magnetic field fluctuations (i.e. the waves) in velocity units $\ba \equiv \delta \B/\sqrt{4\pi \rho}$. We will work in the comoving frame in which $\delta \bm{u}=\bm{u}$ ($\overline{\bm{u}}=0$). The magnetic field strength is $B=|\B|$ or $\vam=|\va|$, and we will use $\hat{\cdot}$ to denote unit vectors (e.g., $\vah = \va/\vam$). In the discussion of waves, $\p$ denotes the wavevector, which makes 
an angle $\tpb$ to $\va$ ($\cos\tpb = \ph\cdot \vah $). In the discussion of expansion, $a$ will denote 
the plasma's expansion factor (starting from some reference position with $a=1$) and the coordinate system 
is Cartesian $\bm{x}=(x,y,z)$, with $x$ the radial direction (that of the mean flow in the solar wind), $y$ the tangential direction, and $z$ the normal direction.

The basis for all of our discussion is the realisation \cite{Barnes1974} that  
\begin{equation}
    P  = {\rm const.},\quad\rho= {\rm const.},\quad B^2 = {\rm const.},\quad \delta \bm{u} = \pm\ba \label{eq: nl aw solution}
\end{equation}
is a nonlinear solution to the compressible MHD equations, which propagates along the mean field  at  the Alfv\'en speed. More generally, the solution \eqref{eq: nl aw solution}
is valid even for  the Kinetic MHD equations that describe large-scale collisionless dynamics \cite{Kulsrud1983,Snyder1997},  if $P$ is replaced with separate constraints on the perpendicular and
parallel pressure and $\va$ is modified to account for any mean pressure anisotropy (in fact, it is valid under even more general conditions  than this; see Ref.~\onlinecite{Barnes1971}).
 Here $\ba$ can be of arbitrary amplitude
compared to $\va$, but $B =|\mB + \delta \B| $ (or $|\va+\ba|={\rm const.}$) must include both  the mean-field and fluctuation contributions. 
Such solutions are the  natural  generalization of Alfv\'enic fluctuations to nonlinear amplitudes \cite{Barnes1974} and are often referred to as spherically polarized waves.

\subsection{One-dimensional nonlinear Alfv\'en waves and field reversals}\label{sub: dbprl derivations}

Although Eq.~\eqref{eq: nl aw solution} is valid for 
general three-dimensional $\ba$, we now specialize 
to 1-D solutions that vary only along the direction $\ph$, so that $\ba(\bm{x})=\ba(\bm{p}\cdot\bm{x})=\ba(\lambda)$.
The functional form $\ba(\lambda)$ is arbitrary, provided that it satisfies the constraints $\nabla\cdot\ba=0$, $\overline{\ba}=0$, and $B={\rm const.}$
Using $\nabla\cdot\ba=\bm{p}\cdot d\ba/d\lambda=0$, which implies $\ph\cdot\ba = 0$ via $\overline{\ba}=0$, we see that $\ba$ has two independent nonzero components. We choose these to be in the $\hat{\bm{n}} = (\ph\times \vah)/|\ph\times \vah|$ direction (the perturbation direction of a linear Alfv\'en wave) and the $\hat{\bm{m}} = (\ph\times \hat{\bm{n}})/|\ph\times \hat{\bm{n}}|$ direction. 
Defining $b_{n}=\hat{\bm{n}}\cdot\ba$ and $b_{m}=\hat{\bm{m}}\cdot\ba$, we see that the 
parallel field perturbation  --- that which can lead to switchbacks --- is 
\begin{equation}
b_{\|} = \vah\cdot \ba = \vah\cdot\hat{\bm{m}}\, b_{m} = b_{m}\sin\tpb.\label{eq: bprl defn}
\end{equation}

We are interested in relating $b_{\|}/\vam$, the relative parallel-field perturbation, to 
the wave's amplitude 
\begin{equation}
A \equiv \frac{|\ba|}{\vam} = \frac{1}{\vam}\sqrt{b_{n}^{2} + b_{m}^{2}}.\label{eq: A defn}
\end{equation}
A simple argument from \alfreds\ (see also Refs.~\onlinecite{Barnes1974,Vasquez1998})
goes as follows. First, note that the constant-$B$ condition,
\begin{equation}
\frac{B^{2}}{{4\pi\rho}} = v_{{\rm A}p}^{2} + b_{n}^{2} + (b_{m}+ v_{{\rm A}m})^{2}\label{eq: const. b condition}
\end{equation}
(where $v_{{\rm A}p} = \ph\cdot\va$ and   $v_{{\rm A}m} = \hat{\bm{m}}\cdot\va$), 
implies $A^{2}/\sin^{2}\tpb + 1 + 2b_{m}/v_{{\rm A}m}={\rm const.}$ on dividing by $v_{{\rm A}m}^{2}$. 
Taking the spatial average and multplying by 
$\sin^{2}\tpb$ fixes the constant, thus allowing one to solve for $b_{m}$ to find 
${b_{\|}}/{\vam} = (\overline{A^{2}} - A^{2})/2$.
This shows that for $A\ll1$, $b_{\|}/\vam$ scales with the wave amplitude squared. 
On the other hand, the amplitude definition \eqref{eq: A defn} shows that $b_{m}/\vam\lesssim A$, 
so that for $A\gtrsim1$, $b_{\|}/\vam\lesssim A\sin\tpb$, with approximate 
equality for $b_{m}\approx b_{n}$. 
Combined, these constraints  give
\begin{equation}
    \frac{b_\|}{\vam} \sim \min\left\{A^2, A\sin\tpb\right\},\label{eq: dbprl}
\end{equation}
which we find is very well satisfied by constant-$B$ 1-D solutions. 
Note that in using \cref{eq: dbprl}, $A$ and $b_{\|}$ are simply numbers rather than functions of $\lambda$  (i.e., $A$ is loosely equated with its spatial average, $A\sim \mA$), a distinction that should be clear from context 
in the discussion below.
We see that large-amplitude waves, with $A\gtrsim \sin\tpb$, preferentially form switchbacks ($b_{\|}\gtrsim \vam$) when they are oblique ($\sin\tpb\approx1$). Unsurprisingly, waves with $A\ll\sin\tpb$ can
never form switchbacks, because, given that $\sin\tpb<1$, the $A^2$ scaling only ever applies for $A\lesssim1$ (implying $b_{\|}\lesssim \vam$).

The result \eqref{eq: dbprl} can be simply explained intuitively as follows: 
field perturbations are confined to be perpendicular to $\bm{p}$ by $\nabla\cdot\B=0$, so $b_{\|}$ perturbations are necessarily 
small when $\bm{p}$ and $\va$ are nearly parallel; additionally,
in large-amplitude waves, $b_{n} $ causes large
variation in $B$, which must be compensated by a similar-magnitude $b_{m}$ component. Thus, large switchbacks result from large-amplitude, oblique nonlinear Alfv\'en waves. Importantly, as discussed by \alfreds\, this provides
a simple explanation for the observed radial elongation of switchbacks \cite{Horbury2020,Laker2021}: in a random magnetic field with power
spread across a wide range of wavenumbers, only preferentially perpendicular (radially elongated) structures generate significant $b_\|$, even for $A\gtrsim1$.

\subsection{Constant-$B$ wave solutions }\label{sub: forming constant B solns}

A question that naturally arises, and one that will be important for our analysis 
later in this manuscript, is how to form constant-$B$ solutions. In general, this 
is simple in one dimension for small-amplitude perturbations, but not for larger amplitudes, and not in 
two or three dimensions (see, e.g., Refs.~\onlinecite{Roberts2012,Primavera2019,Valentini2019} and \zades). 
In one dimension, one can simply arbitrarily 
specify the functional form of $b_{n}(\lambda)$, then solve \cref{eq: const. b condition}
for $b_{m}$:
\begin{equation}
b_{m} = -v_{{\rm A}m} + \sqrt{\frac{B^{2}}{4\pi\rho} - v_{{\rm A}p}^{2}- b_{n}^{2}}.\label{eq: bm solution for ICs}
\end{equation}
In doing so, one must enforce $\overline{b_{m}}=0$ (otherwise $b_{m}$ would contribute to $\va$), which
is used as a constraint to solve  for  the unknown constant field magnitude $B^{2}$.
But, \cref{eq: bm solution for ICs} does not always have real solutions for an arbitrary choice of $b_{n}$, 
which leads to an effective amplitude limit on constructing such a wave for the chosen functional form of $b_{n}$. For
example, Ref.~\onlinecite{Barnes1974} show that if $b_{n}$ is chosen as sinusoidal, with $b_{n}=A_{n}\sin(k\lambda)$, 
solutions on a single branch of the square root exist only for  $A_{n}< (\pi/2) \sin\tpb $.

However, this amplitude limit is artificial: it arises because
the constant-$B$ requirement constrains the functional form of both $b_{n}$ and $b_{m}$ simultaneously, meaning $b_{n}$ cannot 
be chosen arbitrarily. 
If $b_{n}$ is chosen correctly, it is perfectly possible to form smooth constant-$B$ waves 
of arbitrary amplitude for arbitrary $\tpb$. Below, we discuss how to do this by starting
a with lower-amplitude wave, for which one can solve \cref{eq: bm solution for ICs}, then growing 
it via expansion using the asymptotic theory of \alfreds\ (see \cref{eq: alfreds b eqn}). 
Such a process is also physical: Refs.~\onlinecite{Hollweg1974,Barnes1974} showed 
that, no matter how large it becomes, the average amplitude $\mA$ of a spherically polarized 1-D wave grows due to plasma expansion in exactly 
the same way as does a linear  ($A\ll1$) Alfv\'en wave. Thus, starting from small amplitudes
 in the lower solar atmosphere, and neglecting the influence of turbulence and parametric decay, waves can 
 in principle grow to $A\gtrsim1$, so long as both $b_{n}$ and $b_{m}$ change shape in order to 
 allow a consistent solution with constant $B$.

\subsection{Wave evolution and growth with expansion}\label{sub: waves with expansion}
 
 \alfreds\ consider 1-D wave evolution in the MHD expanding-box model  (EBM) \cite{Grappin1993},
 which describes the evolution of a small patch of plasma in a spherically expanding wind with constant velocity $U$. 
 We do not reproduce the full equations here (see \zades), but just note that the model is 
 parameterised by the expansion parameter $a$, which starts at $a=1$ at some reference radius $R_{0}$ and
 evolves as $a=1+\dot{a}t$, where $\dot{a} /a= U/R = U/(R_{0}+U t)$ is the expansion rate ($\dot{a} = U/R_{0}$ is constant). The perpendicular dimensions of a plasma parcel
 scale ${\propto} a$, due to the spherical expansion, while the parallel dimension remains constant. Thus, 
 the wavevector  scales as $\p = (p_{x},p_{y},p_{z}) = (p_{x0},a^{-1}p_{y0}, a^{-1}p_{z0})$, where $\p_{0}=\p(a=1)$ (with components $p_{i0}$). Due to magnetic flux and mass conservation, the Alfv\'en speed evolves with expansion
 according to $\va = (v_{{\rm A}x}, v_{{\rm A}y}, v_{{\rm A}z})= (a^{-1}v_{{\rm A}x0}, v_{{\rm A}y0}, v_{{\rm A}z0}) $ 
 where $\bm{v}_{{\rm A}0}=\va(a=1)$. The different scalings of the radial and perpendicular components 
 of $\va$ cause the mean-field direction to rotate with $a$ and is the manifestation of the
 Parker spiral within this simplified model. High frequency waves, those with frequency $\omega_{\rm A}\gg\dot{a}/a$ (termed the WKB regime), 
 change in amplitude according to $\overline{|\ba|}\propto a^{-1/2}$, independently
 of the wave propagation direction (see App.~\ref{app: zpx sclaing}). Combined with the scaling of $\va$, this leads
 to the well-known result that $\mA\propto a^{1/2}$ in a radial background field, which, as mentioned above, 
 applies for both linear and nonlinear waves ($A\gtrsim 1$).
 Note that, by assuming a constant $U$, the EBM applies only to radii  where $U\gg\vam$, outside the Alfv\'en radius $R_{\rm A}$; inner regions with $R\lesssim R_{\rm A}$ exhibit some very important differences as concerns switchback formation, but are  more complex to study in detail. See, e.g.,  \S II \!D of \zades\  for further discussion. 
 
 \alfreds\ derive a simple equation that captures the slow growth and evolution of $\ba(\lambda)$ due to constant expansion. The method involves using an asymptotic expansion in $\epsilon = (\dot{a}/a)/(\p\cdot\va)$, which is the ratio of the expansion rate to the wave frequency $\omega_{A}=\p\cdot\va$; $\epsilon\ll1$ implies the waves are in the WKB regime (note that $\epsilon$ remains constant during expansion because $\omega_{A}\propto a^{-1}$).
 Averaging over the fast wave motion, they obtain
 \begin{equation}
\frac{\partial \ba}{\partial t} + \frac{\dot{a}}{2a}\left(\ba + {2\ph \hat{p}_{x}}b_{x}\right) + \frac{\epsilon}{2}\frac{\partial}{\partial\lambda} \left[\p\cdot \bm{u}_{1}\, (\bm{v}_{{\rm AT}} + \ba)\right],\label{eq: alfreds b eqn}
\end{equation}
where $\bm{v}_{{\rm AT}}\equiv\va -\ph\,\ph\cdot\va$ is the transverse part of the mean field. The evolution is closed by specifying $\p\cdot \bm{u}_{1}$, which is the (higher-order) compressive flow that is responsible for changing 
the shape of $\ba$ to maintain constant $B$; it is given by 
\begin{equation}
\p\cdot\frac{\partial \bm{u}_{1} }{\partial\lambda}= -\omega_{A}\frac{2b_{x}(v_{{\rm A}x}+ \hat{p}_{x}\ph\cdot\va) - \ba\cdot\va }{|\bm{v}_{{\rm AT}}+\ba|^{2}}.\label{eq: alfreds u1 eqn}
\end{equation}
Wave amplitude growth is contained in the second term in \cref{eq: alfreds b eqn} (coupled with the scaling of $\va$), 
which implies $\overline{|\ba|}\propto a^{-1/2}$ as expected. 
In addition,  \alfreds\ derive equations for the higher-order density and  $B^{2}$ fluctuations that are driven as part of
this process, but these will not feature in our discussion.

Later in this work, we examine the properties of solutions to \cref{eq: alfreds b eqn,eq: alfreds u1 eqn} by means of its 
numerical solution. For this, we use a standard Fourier pseudospectral method with fourth- and fifth-order Runga-Kutta timestepping and 1024 grid points in $\lambda$. We also provide some cursory comparisons of the solutions \cref{eq: alfreds b eqn,eq: alfreds u1 eqn}
with true MHD solutions in App.~\ref{app: comparison to MHD}, finding mostly good agreement except for a specific  case
where \cref{eq: alfreds b eqn} seems to break down (it fails to maintain constant $B$).

Finally, it is worth noting that \cref{eq: alfreds b eqn,eq: alfreds u1 eqn}  provide a convenient and practical  method to construct 
large-amplitude Alfv\'enic solutions when the method described above (\cref{sub: forming constant B solns}) fails at large $A$.
One simply starts with a chosen form of $\ba(\lambda)$ at smaller $A$, then evolves it according to  \cref{eq: alfreds b eqn,eq: alfreds u1 eqn} to reach any desired amplitude.
This process demonstrates that in one dimension at least, the apparent limits on $A$ for some chosen form of $b_{n}$ are artificial; in
all choices for $b_{n}$ that we have tried, waves 
 grow to arbitrary amplitude without forming discontinuities.

\section{Switchback formation due to  expansion}\label{sec: SB formation due to expansion}

\begin{figure}
\centering
\includegraphics[width=1.0\columnwidth]{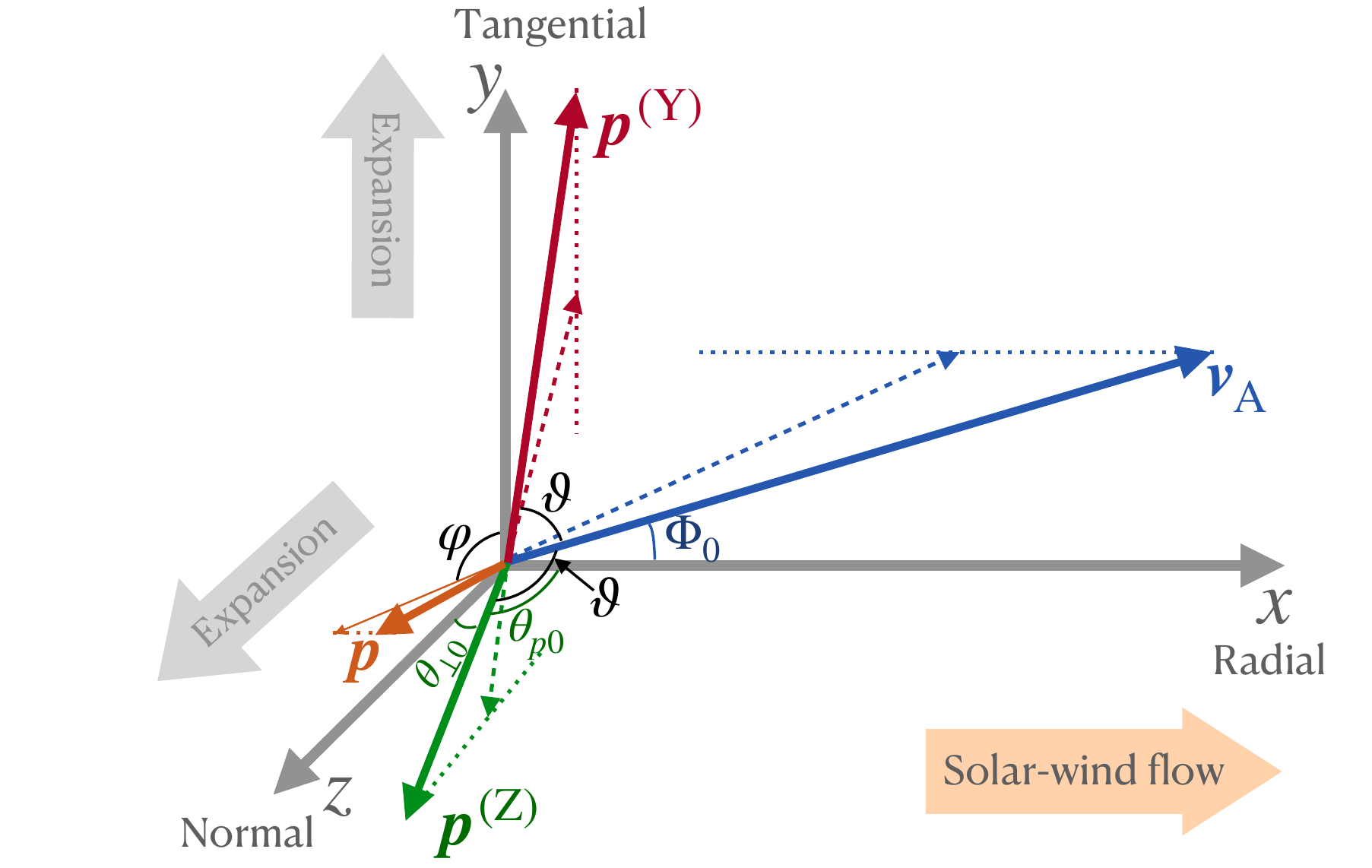}
\caption{The geometry and notation used to describe 1-D 
wave evolution from expansion in a Parker spiral. The mean field $\va$ (blue) and wavevector $\p$ (red, green, and orange) 
evolve according to the EBM scalings described in \cref{sub: waves with expansion}, as shown by the dashed 
arrows and dotted lines. Wave properties
differ significantly depending on the direction of $\p$ compared to the plane of the Parker spiral, 
which we parametrize with  angle $\varphi$, with the extreme cases $\ph^{\rm (Y)}$ and $\ph^{\rm (Z)}$ ($\varphi = 0$ or $\pi$, and $\varphi =\pm \pi/2$, respectively) 
plotted in \cref{fig: theory ps sbs}. The angle between $\va$ and $\p$ plays a key role in determining switchback 
properties and is denoted by $\tpb$.}
\label{fig: geometry}
\end{figure}

In this section, we study the evolution of 1-D Alfv\'en waves in an expanding plasma using
the simple analytic scalings of $b_{\|}$ and $A$ outlined above (\cref{sub: waves with expansion}).
Throughout, we will heuristically equate $b_{\|}/\vam$  with the ``prevalence'' of 
switchbacks, estimating how the this evolves due to 
expansion using \cref{eq: dbprl}. Correspondingly, we will regard $A$ as a number (evolving with $a$) rather
than a function of $\lambda$  throughout this section. We will consider the wave to exhibit strong switchbacks once  $b_{\|}/\vam\gtrsim 1$, 
which (as we will see below) is actually a rather conservative estimate in some regimes because the 
volume filling fraction of large-$b_{\|}$ deviations can evolve to be quite small.

While we will find that the influence of the Parker spiral on switchbacks can be quite dramatic, it is worth cautioning
the detailed results of this section (e.g., the exact form of $b_{\|}(a)$) may be of limited applicability to a real plasma. In particular, they apply only in the absence of turbulence, which 
both causes energy decay (thus changing the amplitude scalings) and acts to populate different wavenumbers $\p$ 
(thus invalidating the assumed wave angle $\tpb$ evolution). Nonetheless, we 
will be able to draw some more general conclusions, which seem to match key results from 3-D nonlinear  EBM simulations (\zades). These issues will be appraised in detail in \cref{sec: discussion}; for now, we simply
take as given the applicability of 1-D wave scalings and examine their consequences.

\subsection{Switchback growth in a radial field}\label{sub: sb growth without ps}

Before considering the evolution of waves in a Parker spiral, it is 
helpful to examine the radial-field case to better understand its important features. 
In this case, $\va = a^{-1}v_{{\rm A}x}\hat{\bm{x}} $, so $A = A_{0} a^{1/2}$, where $A_{0}$
is the initial amplitude at $a=1$. We take $\bm{p}(a)=p_0(\cos\tpi,0,a^{-1}\sin\tpi)$, where 
$\tpi$ is the initial angle between the radial direction and the wave, which we have arbitrarily taken 
to lie in the $x,z$ plane. We also define $ \tp\equiv \tan^{-1}(p_{z}/p_{x}) =\tan^{-1}(a^{-1}\tan\tpi) $ as 
the $a$-dependent evolution of this angle. To obtain simple, physically intuitive results, 
we imagine starting
with a nearly perpendicular wave $\tppi =\pi/2-\tpi\ll1$, and
treat  $\tppi\ll1$ as a small expansion parameter (see \cref{fig: geometry}).

Using  $\tpb=\tp$ in Eq.~\eqref{eq: dbprl}  then expanding in $\tppi\ll1$ gives 
\begin{equation}
\sin^{2}\tpb = \frac{1}{1+a^{2}\tan^{2}\tppi}\approx \frac{1}{1+a^2 \tppi^2},
\end{equation}
 which shows $\sin\tpb\simeq1$ for $a\ll 1/\tppi $ ($\tp\gtrsim1$; mostly oblique propagation) and $\sin\tpb\propto1/a$ for $a\gg 1/\tppi$
  ($\tp\lesssim 1$; mostly parallel propagation). Thus, combined with the continual increase of  $A\propto a^{1/2}$, 
we see that $b_{\|}/\vam$ grows as $b_{\|}/\vam\propto a$ for $a\ll 1/\tppi$, or as $b_{\|}/\vam\propto a^{1/2}$ if  $b_{\|}/\vam\gtrsim1$. Then, it reaches a maximum at $a\approx1/\tppi$, before decreasing as 
$b_\|/\vam\propto a^{-1/2}$ for $a\gg 1/\tppi$ even though $A$ continues to increase.
The cause of the transition between these regimes is simply the transition
from  oblique  ($\tpb=\tp\gtrsim1$) to parallel propagation ($\tpb\lesssim1$), which necessarily implies small $b_{\|}$ perturbations.
The decrease in $b_{\|}/\vam$ once $\tpb\gtrsim1$ also implies 
that  $A_0$ must satisfy $A_0^2\gtrsim\tppi$
in order to form switchbacks at all -- i.e., for $A_0^2\lesssim\tppi$, $b_{\|}/\vam$ reaches a maximum value $b_{\|}/\vam\lesssim1$ before decreasing again. 

An example $b_{\|}/\vam$ evolution, which involves each of the regimes discussed above, is illustrated in \cref{fig: theory ps sbs} with the thick black lines.

\subsection{Switchback growth in the Parker Spiral}\label{sub: sb growth in ps}

In the presence of a perpendicular component to the mean field (the Parker spiral), 
the scalings described above become more complex and interesting because of non-monotonic behavior of $\sin\tpb$ and $A$. First, let us consider 
the amplitude evolution. We take the Parker spiral to lie in the $x,y$ plane, $\bm{v}_{{\rm A}0} = v_{{\rm A}0}(\cos\tpsi, \sin\tpsi, 0)$, which 
implies that the radial (R), tangential (T), and normal (N) directions correspond to $x$, $y$, and $z$, respectively (see \cref{fig: geometry}). We define $0<\tpsi\ll 1$ as the initial Parker Spiral angle, which 
(like $\tppi$) will be considered a small parameter and used to simplify the results. We also define the $a$-dependent Parker spiral angle $\tps \equiv \tan^{-1}(v_{{\rm A}y}/v_{{\rm A}x}) = \tan^{-1}(a \tan\tpsi) $.
Using  $\va = v_{{\rm A}0}(a^{-1}\cos\tpsi, \sin\tpsi,0)$ and  $\overline{|\ba|}\propto a^{-1/2}$ gives,
\begin{equation}
    A \approx \frac{A_0 a^{-1/2}}{\sqrt{a^{-2}\cos^2\tpsi+\sin^2\tpsi}}\approx \frac{A_0a^{1/2}}{\sqrt{1+a^2\tpsi^2}},\label{eq: A evolution with PS}
\end{equation}
showing that $A  $ grows like the radial-field case, $A\propto a^{1/2}$, for $\tps\lesssim1$ ($a \lesssim 1/\tpsi$), 
but starts decreasing $A\propto a^{-1/2}$ once $\tps\gtrsim1$ ($a\gtrsim 1/\tpsi$) because the mean field decays more slowly than the wave-like perturbations. 

We must now  allow
$\p$ to have components in all three directions to capture the full range
of possible behaviors of $\sin\tpb$. We thus parameterize it with 
$\bm{p}(a)=p_0(\cos\tpi,a^{-1}\sin\tpi \cos\varphi,a^{-1}\sin\tpi \sin\varphi)$, 
so that $\varphi = \pm\pi/2$ corresponds to $\p$ lying in the  plane perpendicular to the Parker spiral mean field, and $\varphi = 0$ or $\pi$ corresponds to 
$\p$ lying in the  same plane as the Parker spiral. We will sometimes describe these cases as $\ph^{{\rm (Z)}}$ and $\ph^{{\rm (Y)}}$, respectively,
as illustrated with the green and red
arrows  in \cref{fig: geometry}.
 We again imagine starting from a highly oblique wave  ($\tppi=\pi/2-\tpi\ll1$, also with $|\tpb|<\pi/2$) and compute 
\begin{equation}
    \sin^2\tpb = 1-\frac{(\bm{p}\cdot\va)^{2}}{|\bm{p}|^{2}|\va|^{2}}\label{eq: sin2 theta for waves}
\end{equation}
as a function of $a$.
The result is that $\sin\tpb$  decreases in the same way as radial-$\va$ case initially, but its evolution starts to  differ markedly as $a$ approaches 
 \begin{equation}
    a_{\tpb_{\rm min}} = \sqrt{\cot\tppi\cot\tpsi} \approx \frac{1}{\sqrt{\tppi\tpsi}},\label{eq: amin parker spiral}
\end{equation} 
at which point 
 $\sin\tpb$ reaches a local minimum
 \begin{align}
\sin^2\tpb_{\rm \min} &=  1- \frac{(\sin\tpsi\cos\tppi\cos\varphi + \cos\tpsi\sin\tppi)^{2}}{\sin^{2}(\tpsi + \tppi)}\nonumber\\
& \approx 1-\frac{(\tpsi \cos\varphi + \tppi)^{2}}{(\tppi+\tpsi)^2}.\label{eq: sinmin parker spiral}
\end{align} 
  From this point, unlike the radial case, $\sin\tpb$ starts increasing again back towards oblique propagation, because $\va$ rotates towards the perpendicular direction.

\begin{figure}
    \centering
    \includegraphics[width=0.99\columnwidth]{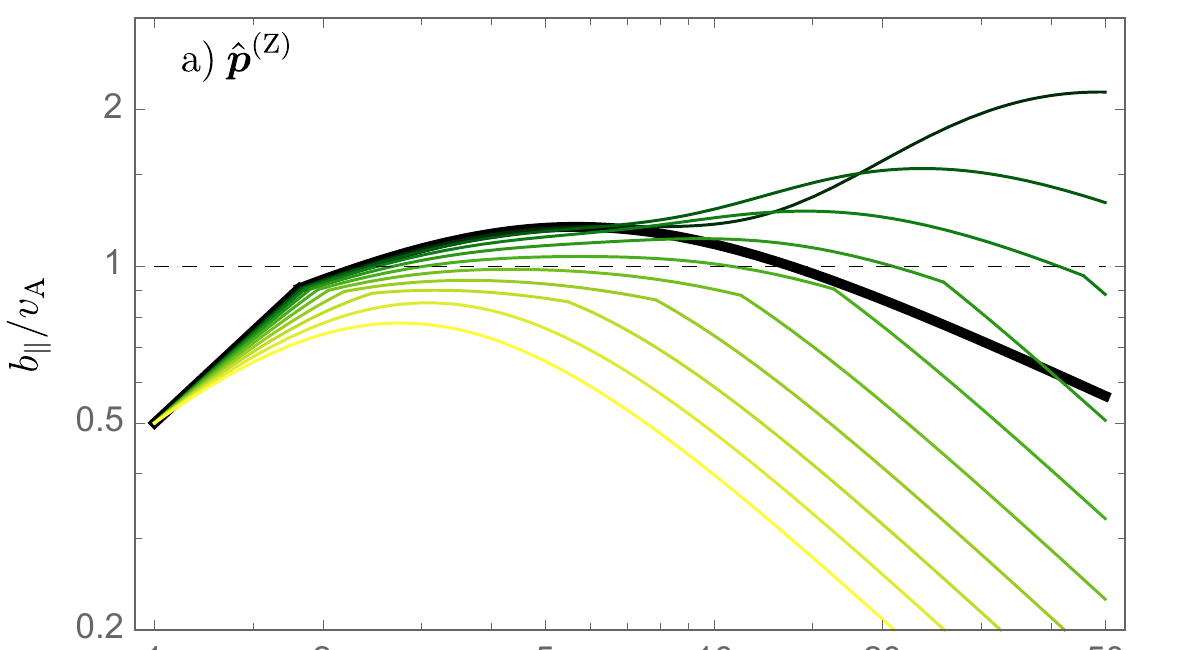}\\\vspace{-0.2cm}~\includegraphics[width=1\columnwidth]{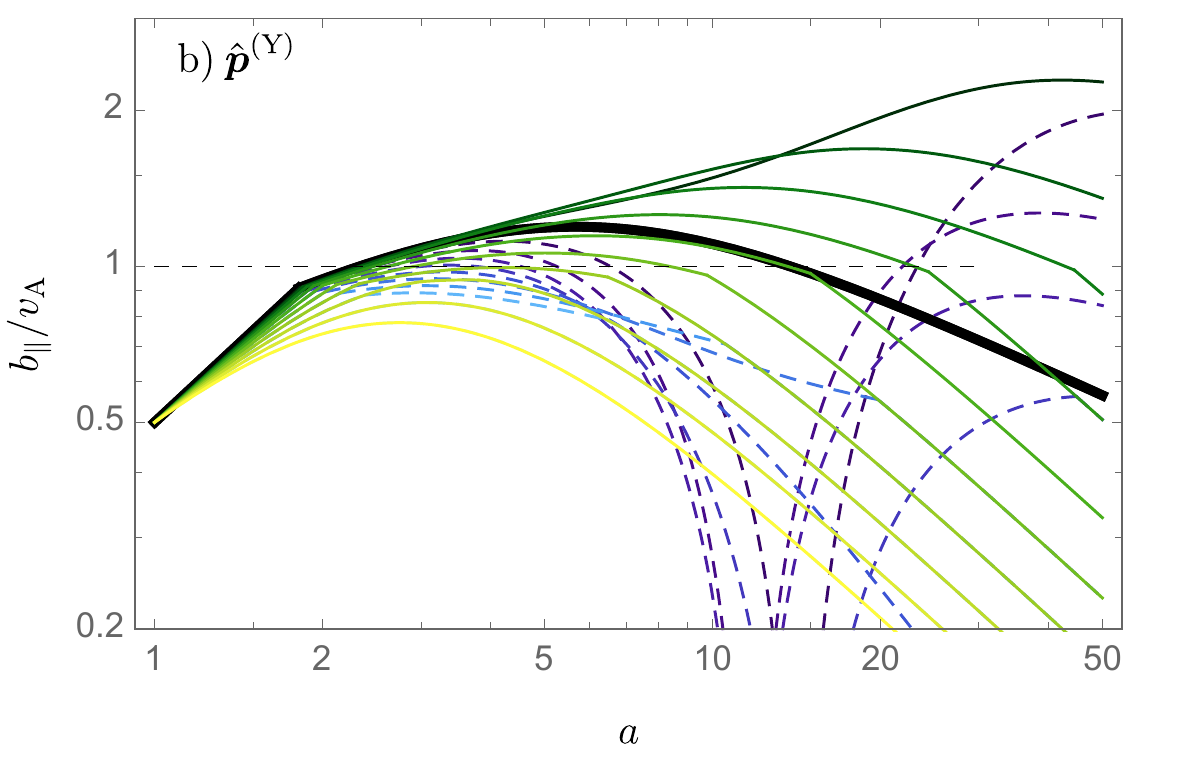}
    \caption{Evolution of the parallel-field perturbation, roughly equatable with
    the switchback prevalence, computed from Eq.~\eqref{eq: dbprl} using the EBM scalings
    of $A$ and $\tpb$. In all cases we start from $\tpb=\pm80^\circ$ and $A_0$=0.8; thick black lines
    show the radial-$\va$ case ($\tpsi=0$), and colored lines show that with a Parker Spiral with $\tpsi=2^\circ, 4^\circ, \dots, 20^\circ$ from dark green to light yellow, or dark blue to light blue. 
    Panel (a) shows $\varphi=\pm\pi/2$, where $\bm{p}$ lies in the $x,z$ plane  ($\hat{\bm{p}}^{{\rm (Z)}}$ in \cref{fig: geometry}), while panel (b) shows $\varphi=\pi$ in solid green-yellow lines and $\varphi=0$ in dashed blue lines (in both these cases $\bm{p}$ lies in the $x,y$ plane; $\hat{\bm{p}}^{{\rm (Y)}}$ in \cref{fig: geometry}). For $\varphi=0$, the wave passes through purely parallel propagation at 
    $a=a_{\tpb_{{\rm min}}}$ (\cref{eq: amin parker spiral}), implying  
   $b_{\|}/\vam=0$.}
    \label{fig: theory ps sbs}
\end{figure}

  We illustrate the effect of these features on the  evolution of  \cref{eq: dbprl} in \cref{fig: theory ps sbs}, using \cref{eq: A evolution with PS,eq: sin2 theta for waves} for $A$ and $\sin\tpb$. 
All curves have the same initial $\tpb=80^{\circ}$, and the different colors show  
   a variety of initial Parker spiral angles $\tpsi$. 
   Note that $\tpi$ must be adjusted to keep  fixed initial $\tpb $ while varying $\tpsi$, which implies $\tpi = \cos^{-1}(\cos\tpb/\cos\tpsi)$ for $\varphi=\pm\pi/2$, or  $\tpi=\tpb\cos\varphi +\tpsi$ for $\varphi = 0$ or $\pi$.
   We plot the waves with $\varphi =\pm\pi/2$ ($\ph^{{\rm (Z)}}$) and with $\varphi = 0,\pi$ ($\ph^{{\rm (Y)}}$) separately in the two panels, because their evolution differs significantly and this 
will suggest important conclusions about switchback properties. 
 In order to understand the illustrated behavior, let us first 
 consider the importance of the two angles $\tpsi$ and $\tppi$ (controlling the Parker spiral and the wave's obliquity, respectively), then consider the wave's orientation $\varphi$. 

For $\tppi\gtrsim\tpsi$ --- i.e., when the Parker Spiral makes a smaller angle to the 
radial than the wavevector makes to the perpendicular --- $a_{\tpb_{\rm min}}<1/\Phi_0$. This means $\sin\tpb$ reaches its minimum and starts increasing again {before} $A$ starts decreasing at $\tps=45^{\circ}$ ($a\sim 1/\tpsi$; see \cref{eq: A evolution with PS}), but after the maximum of $b_{\|}/\vam$ for the radial-$\va$ waves at $a\sim 1/\tppi$. This explains the inflection points seen in the green curves in \cref{fig: theory ps sbs}: $\sin\tpb$ decreases significantly below unity by $a\sim 1/\tppi$ when $\tp\lesssim1$,  then reaches its minimum at $a\sim 1/\sqrt{\tppi\tpsi}$ causing $b_{\|}/\vam$ to start increasing again; but then $A$ itself starts decreasing at $a\sim 1/\tpsi$ causing $b_{\|}/\vam$ to decrease. Thus, in this regime of modestly oblique waves $\tppi\gtrsim\tpsi$, switchbacks usually
form significantly more efficiently than in the 
radial-$\va$ case, because the waves evolve to become more oblique after $\ph$ rotates to be mostly radial, but before the Parker spiral rotates past $\simeq 45^{\circ}$. 
In contrast, in the opposite regime of a large Parker spiral  with $\tppi\lesssim\tpsi$ (yellow curves in \cref{fig: theory ps sbs}), 
$A$ starts decreasing, at $a\approx 1/\tpsi \lesssim a_{\tpb_{\rm min}}$, before the minimum in $\sin\tpb$. Because $\sin\tpb_{\rm min}\approx 1$  in this regime unless $\cos\varphi\approx1$ (see \cref{eq: sinmin parker spiral}),   $b_\|/\vam \sim \min(A,A^{2})$, which  simply peaks when $\tps\sim1$ then decreases again,  making for inefficient switchback formation even though the wave remains oblique at all times. In both regimes, the decrease $A\propto a^{-1/2}$ always wins out 
and causes $b_{\|}/\vam $ to decrease as $b_{\|}/\vam \propto a^{-1}$  once $\Phi\gtrsim 1$ and $A\lesssim1$, which is faster than in the radial-$\va$ case.

As seen by the comparison between panels (a) and (b) of \cref{fig: theory ps sbs}, the wave's direction $\varphi$ is also key in determining its evolution.
For $\cos\varphi\neq1$, as applies to $\ph^{{\rm (Z)}}$ waves or when $\ph$
and $\va$ lie in the same plane but different quadrants ($\varphi=\pi$), 
the evolution occurs broadly as described above, with $\varphi=\pi$ having a 
modestly larger $\sin\tpb_{\rm min}$ and thus slightly larger $b_{\|}/\vam$ for intermediate times (c.f. green-yellow curves in panels (a) and (b)). However, 
for $\varphi=0$, $\ph$ and $\va$ pass through each other at $a_{\tpb_{\rm min}}$, \emph{viz.,} the wave becomes perfectly parallel with  $\tpb=0$ (this effect 
can be seen clearly by following the dotted red and blue lines in \cref{fig: geometry}). At this point, $b_{\|}/\vam$ must go to zero, and $\tpb$ flips sign.
Although $\sin\tpb$ then increases rather rapidly in the opposite direction, we see that for most of these wave's evolution, they produce only small $b_{\|}/\vam$ fluctuations. Thus, the Parker spiral can create strong differences between switchback properties, depending on the 
direction of fastest variation ($\ph$) of the wave or structure in question.

Finally, we note that in order for any physics related to $\sin\tpb$ to be relevant, the maximum of $b_\|/\vam$ should occur when $A^2\gtrsim A\sin\tpb$ (see Eq.~\eqref{eq: dbprl}). Because the maximum of $b_\|/\vam$ occurs together with the maximum of $A$, which is at $a\sim 1/\tpsi$, we see that for  initial wave amplitudes $A_0^{2}\lesssim \Phi_{0}$,  $b_\|/\vam\sim A^2$ for all $a$ and there are no significant switchbacks ($b_{\|}/\vam \lesssim1$).

\begin{figure}
\centering
\includegraphics[width=1.0\columnwidth]{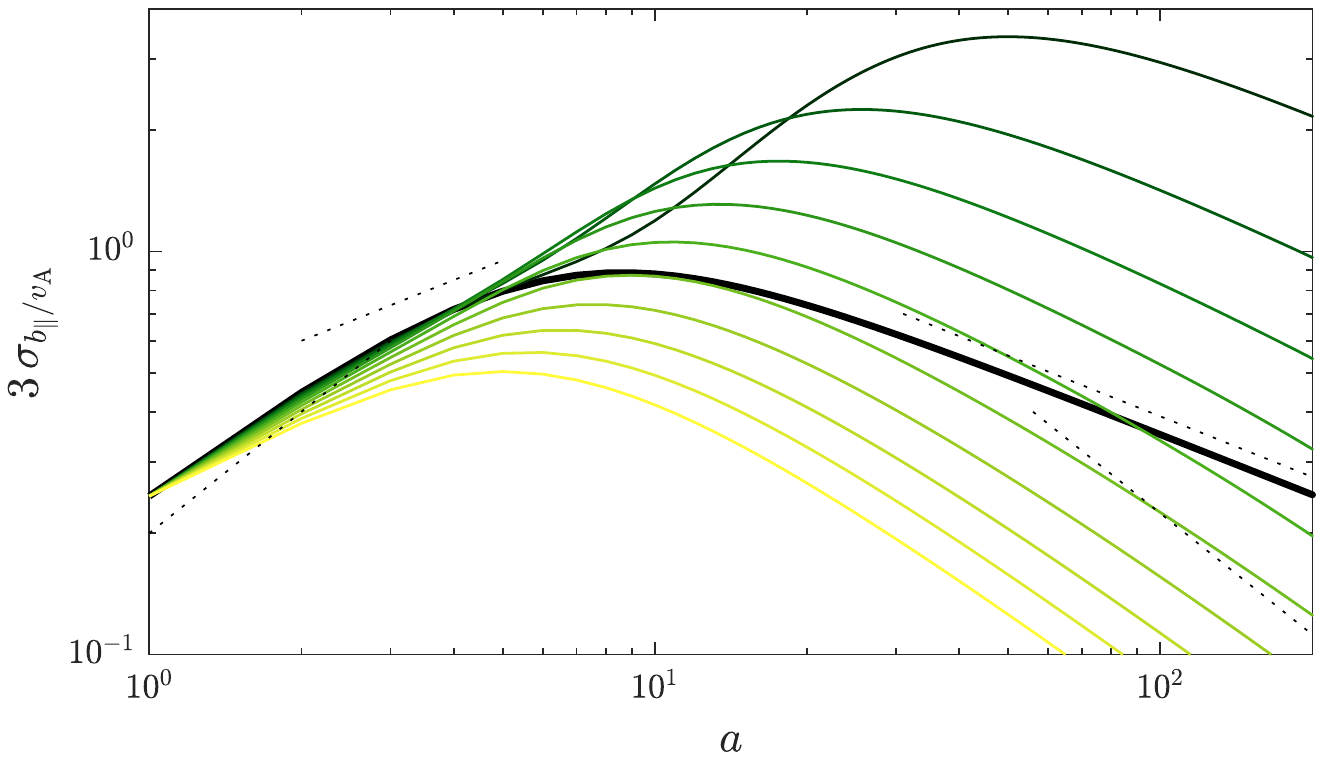}
\caption{Measured switchback prevalence computed from three times the standard deviation of $b_{\|}/\vam$, $\sigma_{b_{\|}/\vam}= [\overline{(b_{\|}-\overline{b_{\|}})^{2}}]^{1/2}/\vam$,
for 1-D waves evolved according to \cref{eq: alfreds b eqn,eq: alfreds u1 eqn}. The
color scheme is the same as \cref{fig: theory ps sbs}, and we take $\bm{p}$ to lie in the $x,z$ plane ($\varphi= \pi/2$). 
The dashed lines show the scalings, from left to right, $a^{1}$, $a^{1/2}$, $a^{-1/2}$, and $a^{-1}$, which 
are the theoretical expectations for the different regimes discussed in \cref{sub: sb growth in ps}. Clearly, aside from 
a modest vertical offset, which arises from the ambiguity of the definition of $b_{\|}/\vam$ in \cref{eq: dbprl}, the 
simple computations of \cref{fig: theory ps sbs} give an extremely good match to the behavior of real 1-D waves (note the extended horizontal axis compared to \cref{fig: ps sbs from waves}).  }
\label{fig: ps sbs from waves}
\end{figure}

\subsubsection{Comparison to wave solutions}\label{subsubs: comparison to wave solutions}

The above arguments and \cref{fig: theory ps sbs} are based purely on
\cref{eq: dbprl}. How well do these estimates compare to true expanding wave solutions?
To test this, we solve \cref{eq: alfreds b eqn,eq: alfreds u1 eqn} with period boundary conditions, 
starting from random initial conditions constructed from the first $10$ Fourier 
modes with $\tpb=80^{\circ}$ and $\varphi=\pi/2$, and  varying $\tpsi$ in order to 
match the computations \cref{fig: theory ps sbs}(a). Results are
shown in \cref{fig: ps sbs from waves}. The definition of $b_{\|}/\vam$ is only valid as a scaling in \cref{eq: dbprl}, so to make a
reasonable comparison we measure the standard deviation of $b_{\|}/\vam$ across the domain, with $b_{\|}(\lambda) =\ba\cdot\vah$. The agreement, both
in the general form and the predicted scalings with $a$ (dotted black lines),
is extremely good, including in features such as the inflection point, which one might have expected to be an  artefact  of the idealised nature of \cref{eq: dbprl}. The spatial form of the solutions themselves are shown in \cref{fig: ps solutions} for $\tpsi=2^{\circ}$ and \cref{fig: no ps solutions} for $\tpsi=0$ (radial $\va$) and will discussed in detail below. In
addition, we show in \cref{app: comparison to MHD} (see \cref{fig: MHD compare radial}) that the solutions of \cref{eq: alfreds b eqn}
match  true 2-D expanding MHD solutions very well.

In  \cref{fig: ps sbs from waves}, we do not consider waves with  $\varphi=0$ or $\pi$, for which $b_{\|}$ estimates are 
shown in \cref{fig: theory ps sbs}(b). The reason for this is that \cref{eq: alfreds b eqn,eq: alfreds u1 eqn}
fail to produce constant-$B$ solutions when $\varphi=0$ or $\pi$. Although the cause for this behavior remains unclear (we speculate
that it results from the more rapid evolution of   $\sin\tpb$ in this geometry), it is important to note that \cref{eq: alfreds b eqn} was derived 
assuming constant $B$, so if this is not satisfied we should not trust its solutions. 
Thus, it is not worthwhile to compare to the predictions 
of \cref{eq: dbprl} and \cref{fig: theory ps sbs}(b) in detail. The consequences of this discrepancy are discussed below (\cref{subsub: py and constant B}) and 
in \cref{app: comparison to MHD} (see \cref{fig: MHD compare Y} for MHD solutions of this geometry).

\subsection{Consequences for the solar wind}

Although we will delay detailed discussion of turbulence and 3-D fields 
until \cref{sec: discussion}, it is helpful to briefly outline some possible observable
consequences of these wave properties for switchbacks in the solar wind.  
The most obvious property from \cref{fig: theory ps sbs} is that a modest Parker spiral, with $\tps\lesssim1$ ($v_{{\rm A}x}>v_{{\rm A}y}$) can
significantly enhance switchback formation. This is because, when $\tppi \gg \tpsi$ (with $\tppi = \pi/2-\tpi \simeq \pi/2-\tpb_{0}$ in most regimes) the simultaneous rotation of $\va$ and $\p$ causes the wave obliquity to increase even when $\tps\ll1$. 
This seems to be the more relevant regime for the solar wind, since we 
 measure $\tps$ to be rather small near  $R_{\rm A}$ (where the EBM becomes applicable), and there are a wider range of wavenumbers with $\tppi\gtrsim \tpsi$ than with $\tppi\lesssim \tpsi$ if $\tpsi\ll1$. Thus, we predict 
more robust growth of switchbacks due to expansion in the presence of a sub-$45^{\circ}$ Parker spiral than not, an observationally
testable prediction that is also seen in the simulations of \zades.

Another interesting conclusion we can draw concerns the directions of 
switchback deflections. All else being equal, waves with $\p$  perpendicular to the plane of the Parker spiral ($\ph^{\rm (Z)}$) generate
more switchbacks than waves with $\p$ in the plane of the Parker spiral ($\ph^{\rm (Y)}$): half of the $\ph^{\rm (Y)}$ waves (those with $\varphi=0$) evolve 
to become purely parallel and cause only small $b_{\|}/\vam$ over a wide portion 
of their evolution. Because these are Alfv\'enic fluctuations, the strongest $\ba$ fluctuation 
lies in the $\hat{\bm{n}}=(\ph\times \vah)/|\ph\times \vah|$  direction, which means that $\ph^{\rm (Z)}$
waves cause large $b_{y}$ fluctuations, and $\ph^{\rm (Y)}$ large $b_{z}$ fluctuations. 
Thus, we expect switchbacks to preferentially 
involve rotations of the field in the plane of the Parker
spiral (the tangential direction), rather than the normal direction.  This 
seems to be observed, at least partially, in the simulations of \zades\ (see their figure 7), and, more clearly in PSP observations \cite{Horbury2020,Laker2022}. 
In essence, this argument is nothing more than the statement that 
for a distribution of waves with wavevectors that are biased towards the radial 
direction (as caused by expansion), wavevectors that lie in the plane
perpendicular to the mean field are more oblique, on average, than those in the plane of the mean field. This interpretation  is explored 
in more detail in \cref{sub: integral estimate} to provide another argument for this general effect.

\subsubsection{The assumption of constant $B$ for  $\ph^{\rm (Y)}$ waves}\label{subsub: py and constant B}

We noted above that when $\varphi =0$ or $\pi$ ($\ph^{\rm (Y)}$), \cref{eq: alfreds b eqn} fails to maintain a constant-$B$
solution as the wave grows. This  raises the obvious question of whether \cref{eq: dbprl}  is valid for such waves, and, if it is not, what will be
the consequences.  In \cref{app: comparison to MHD}, we address this question by
directly comparing solutions of \cref{eq: alfreds b eqn} to   2-D expanding MHD solutions, finding that 
indeed \cref{eq: alfreds b eqn} overpredicts the variation in $B$ compared to MHD for this geometry, although the general
form of the solutions is similar (see \cref{fig: MHD compare Y}).
However, we do still see tentative evidence that, even in MHD,  larger variation in $B$ occurs compared 
to a radial $\va$ or $\varphi=\pm\pi/2$, particularly for $\varphi=\pi$ (which, recall, generates larger $b_{\|}$ than $\varphi=0$).
While a more careful study is needed, if this result holds, it only strengthens our main results from this section, implying
that not only do $\ph^{\rm (Y)}$ waves generate relatively smaller $b_{\|}$ because of the geometry of $\sin\tpb$, they 
also generate larger $B$ fluctuations that will then be more prone to dissipation by other means (e.g.,  kinetic damping or shocks). 
This would only act to enhance the dominance of $b_{y}$ (tangential)  over $b_{z}$ (normal) rotations in switchbacks, 
strengthening our second conclusion above.

\section{The structure of switchbacks in the tangential plane}\label{sec: structure}

\begin{figure*}
    \centering
    \includegraphics[width=1\textwidth]{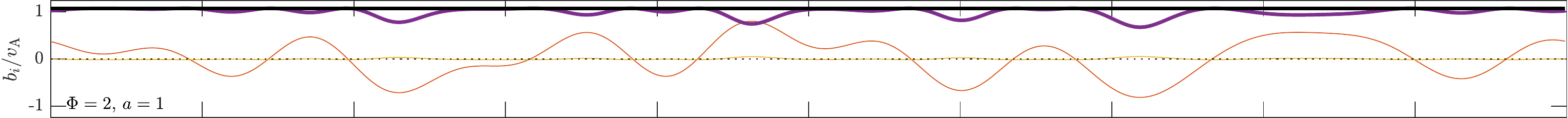}\\
    \includegraphics[width=1\textwidth]{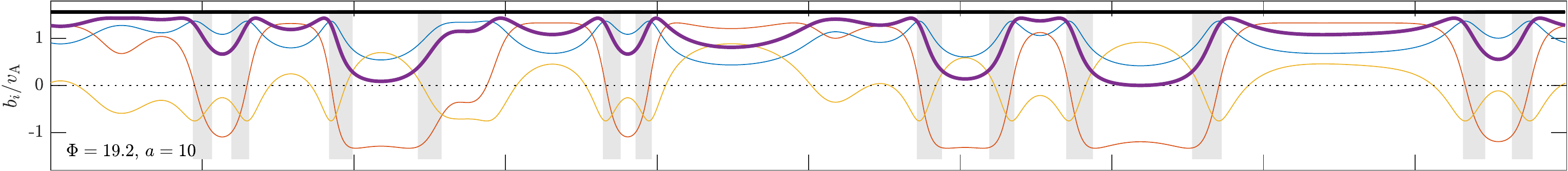}\\
    \includegraphics[width=1\textwidth]{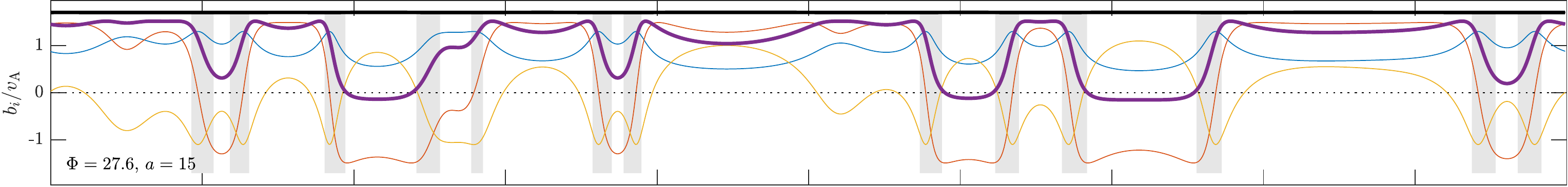}\\
    \includegraphics[width=1\textwidth]{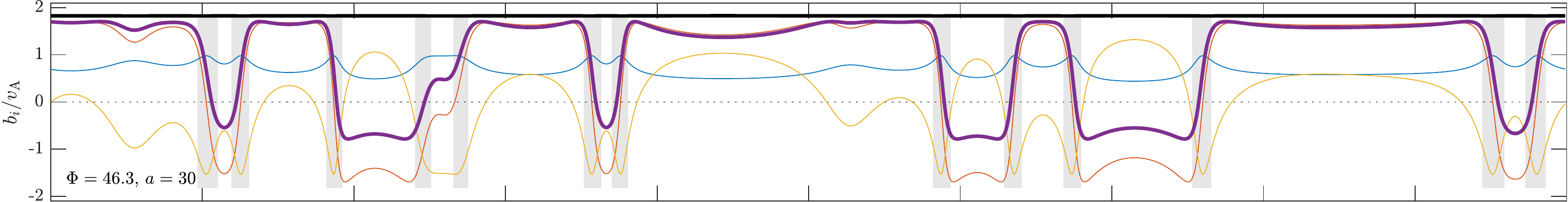}\\
    \includegraphics[width=1\textwidth]{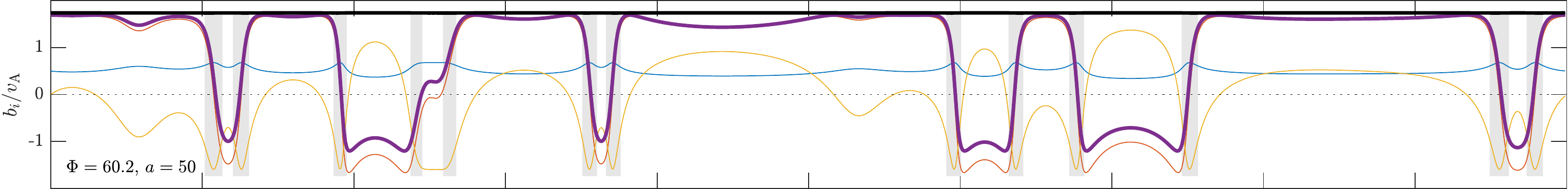}\\
    \includegraphics[width=1.0023\textwidth]{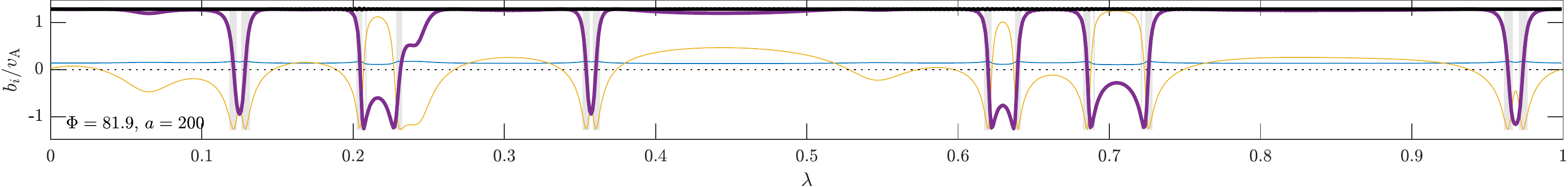}\\
    \caption{An example of how a spherically polarized wave evolves in a background magnetic field with a Parker spiral, computed 
    from \cref{eq: alfreds b eqn,eq: alfreds u1 eqn}. The wave is similar to the
    darkest-green lines in \cref{fig: theory ps sbs}a and \ref{fig: ps sbs from waves}, with $\tpsi=2^{\circ}$,  
    initial $\tpb=80^{\circ}$ (implying $a_{\tpb_{\rm min}}\approx12.7$), and $A_{0}\approx 0.4$ formed from a random collection of the first $5$ Fourier  modes. In each panel, thin blue, red, and yellow lines show $b_{x}+v_{{\rm A}x}$, $b_{y}+v_{{\rm A}y}$, and $b_{z}$, respectively, each normalized to $\vam$. The thick black line shows $(B/\sqrt{4\pi\rho})/\vam$, and the thick purple line shows $b_{\|}/\vam + 1$, which is the normalized 
    total field in the $\vah$ direction. We show panels of increasing $a$, as labelled in the bottom-left corner, with the vertical size of each panel scaled to the current total amplitude of the wave (i.e., the $y$-axis scale remains constant in units of $b_{i}/\vam$). The shaded grey regions in each panel highlight the regions of $b_{\|}$ with particularly
    sharp gradients --- i.e., the switchback deflections --- which we can see are generically associated with increases (rather than decreases) in $b_{x}$. This implies the $\ba$ vector rotates in a tangentially skewed way in switchbacks, towards the $v_{{\rm A}x}$ direction. In addition, we see that switchbacks become more and more intermittent as $\tps$ increases: although the standard deviation of $b_{\|}$ starts decreasing for $a\gtrsim 50$ because $A$ starts decreasing (c.f. \cref{fig: ps sbs from waves}), the relative size of individual deflections actually increases, while their volume filling fraction  decreases. }
    \label{fig: ps solutions}
\end{figure*}

\begin{figure*}
    \centering
    \includegraphics[width=1\textwidth]{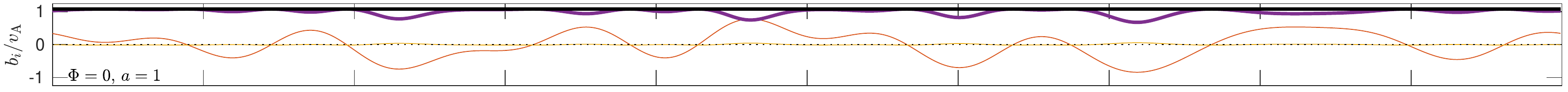}\vspace{-0.1cm}\\
    \includegraphics[width=1\textwidth]{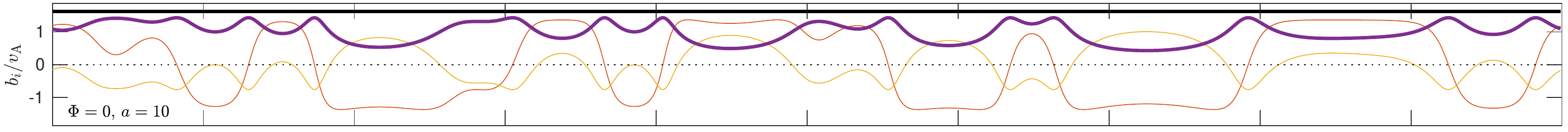}\vspace{-0.1cm}\\
    \includegraphics[width=1\textwidth]{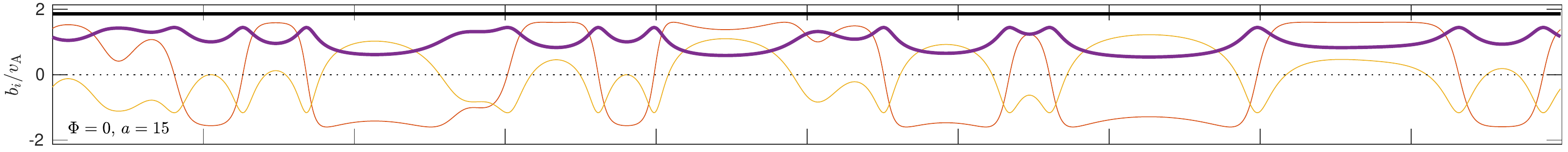}\vspace{-0.1cm}\\
    \includegraphics[width=1.0023\textwidth]{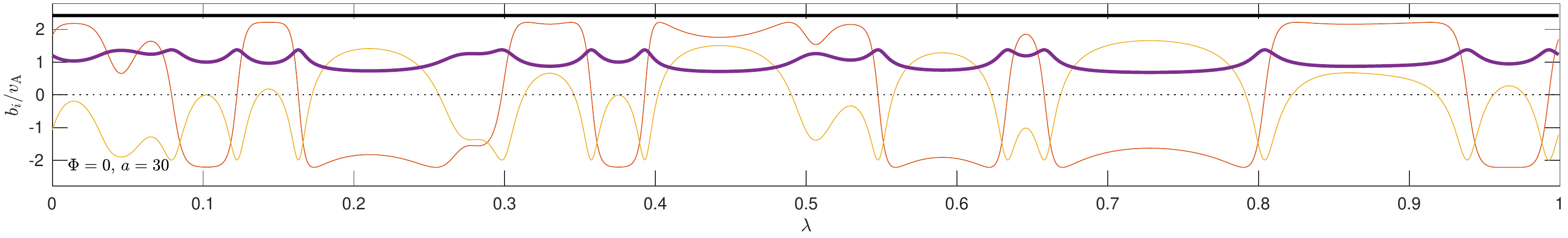}
    \caption{As in \cref{fig: ps solutions}, and with the same $b_{n}$ used for the initial conditions, but with no Parker spiral ($\tps=0$). In this case, 
    $b_{\|}=b_{x}$ so the purple line covers the blue line. The switchback prevalence starts decreasing once  $a\gtrsim10$, even though $A$ continues to increase. Beyond this point, the wave form hardly changes shape as it grows because the mean field 
    makes a relatively small contribution to the total $B$.
     }
    \label{fig: no ps solutions}
\end{figure*}

Above, we saw that the addition of a small Parker spiral can significantly enhance switchback generation, particularly when $\bm{p}$ lies in the $x,z$ plane and $\ba$ is dominated by its $y$ component. 
In this section, we explore the structure of the wave solutions that develop 
under these conditions, governed by the constant-$B$ constraint and the rotating mean field and wavevectors. 
We will show that in the regime where the
Parker spiral significantly modifies  switchbacks --- i.e., as $a$ approaches and exceeds $a_{\tpb_{\rm min}}$ (\cref{eq: amin parker spiral}) --- 
  field rotations are generically ``tangentially skewed'':
the deflection of $\ba$  always causes $b_x$ to increase, rather than decrease. This implies that through 
the switchback, the field deflects around towards the positive radial  ($v_{{\rm A}x}$) direction (then past it), as opposed to the negative radial direction. We
provide a simple proof for why this must occur based on certain
assumptions about the form of such switchbacks and the $\nabla\cdot\bm{b}=0$ and  constant-$B$ constraints. 
For simplicity of notation, we will assume $v_{{\rm A}x}$ points in the $+\hat{\bm{x}}$ direction and $v_{{\rm A}y}$ in the $+\hat{\bm{y}}$ direction; in the case with 
$v_{{\rm A}x}$ in the $-\hat{\bm{x}}$ direction, $\ba$ deflects towards the negative radial direction instead (i.e., still $v_{{\rm A}x}$), but is otherwise identical. We will also assume the wave starts  with $\tppi\gtrsim \tpsi$, because the opposite (large-Parker-spiral) limit with $\tppi\lesssim \tpsi$ was shown above to be ineffective
at generating switchbacks (it also requires extremely perpendicular waves for small  $\tpsi$). Through this 
section we do not consider  waves with $\varphi=0$ or $\pi$ ($\ph^{\rm (Y)}$); this is both because constant-$B$ solutions for such waves are not 
so easily understood (see \cref{subsub: py and constant B}), and because field rotations will be more symmetric in this case anyway (since they 
are predominantly in the normal direction; see \cref{fig: MHD compare Y}).

The evolution of a representative 1-D wave solution with a Parker spiral is shown in \cref{fig: ps solutions}, starting from $\tpb=80^{\circ}$, $\tpsi=2^{\circ}$,
and $\varphi=\pi/2$, to match the parameters of the dark green solutions in Figs.~\ref{fig: theory ps sbs}(a) and \ref{fig: ps sbs from waves}.  The equivalent solution without a Parker spiral, matching the black lines in Figs.~\ref{fig: theory ps sbs}(a) and \ref{fig: ps sbs from waves}, in is shown in \cref{fig: no ps solutions} to provide a
reference for comparison (see also \alfreds\ figure 4). Similar to \cref{fig: ps sbs from waves}, the initial conditions are constructed using \cref{eq: bm solution for ICs} with random amplitudes in the first $5$ Fourier modes for $b_{n}= b_{y}$.
We then evolve these according to \cref{eq: alfreds b eqn,eq: alfreds u1 eqn} with periodic boundary conditions, capturing the change in shape of
$\ba$ needed to keep $B$ constant as the wave grows due to expansion. 
This initial condition, as expected, involves  predominantly $b_{y}$ perturbations (red line). Strong switchbacks (thick purple  lines) develop at later times as expected from \cref{fig: theory ps sbs}a. They  are dominated by sudden changes in the direction of $b_y$ and always involve an increase in $b_{x}$ (blue lines)
through the sharp change in  $b_{\|}$ (grey shaded regions). 
This implies that the magnetic-field lines always rotate  towards the positive radial direction during the field reversal, \emph{viz.,} they are tangentially skewed. Switchbacks do not grow nearly as effectively   in the radial-$\va$ case (\cref{fig: no ps solutions}).

\subsection{A simple proof of switchback tangential skewness }\label{sub: proof of skewness}

To understand why this behavior occurs, let us first consider the regime 
in which it occurs. 
Because $b_{\|} = b_{x}\hat{v}_{{\rm A}x} + b_{y} \hat{v}_{{\rm A}y}$ must decrease below ${\simeq} -\vam$ to form a switchback, it must involve either a large-negative $b_{x}$ or large-negative $b_{y}$ (or both). The former case is simply a standard
radial-$\va$ switchback as explored in \alfreds\ and should not be expected to involve preferential deflections. This
situation will apply even with a Parker spiral when   $a\lesssim 1/\tppi$ (i.e., when $\tp\gtrsim 45^{\circ}$, which also implies $\tps\ll45^{\circ}$), because such waves behave like the radial-$\va$ case anyway (see \cref{sub: sb growth in ps}).
However, the latter case, with a large $b_{y}$ perturbation, is different. As we now
show, it takes over before or around $a\sim a_{\tpb_{\rm min}}$, which is well before $v_{{\rm A}y}\approx v_{{\rm A}x}$ ($\tps\approx 45^{\circ}$; see \cref{eq: amin parker spiral}). To understand why, 
we first note that $b_{y}\approx b_{n}$ and $b_{m}^{2}\approx b_{x}^{2} + b_{z}^{2}$ because $\va$ remains nearly radial, while $\p\cdot\ba=0$ implies \begin{equation}
b_{x} = -\tan\tp b_{z}\label{eq: div b for z sb}
\end{equation}
(this can be clearly observed in the blue and yellow lines in \cref{fig: ps solutions}).
For $a\approx a_{\tpb_{\rm min}}\approx 1/\sqrt{\tppi\tpsi}$ and $\tppi\gg\tpsi$, we thus see $b_{x}\ll b_{z}\approx b_{m}$ implying $b_{x}\approx b_{m}\sqrt{\tpsi/\tppi}$. This shows that at $a\approx a_{\tpb_{\rm min}}$, 
\begin{equation}
 b_{x}\hat{v}_{{\rm A}x}  + b_{y} \hat{v}_{{\rm A}y}  \approx b_{m} \sqrt{\tpsi/\tppi} +  b_{n} \sqrt{\tpsi/\tppi},\label{eq: bprl at amin}
\end{equation}
where we have used  the fact that 
$\hat{v}_{{\rm A}x} \approx 1$ and $\hat{v}_{{\rm A}y}\approx a\tpsi \approx \sqrt{\tpsi/\tppi}$ (because $\tps\ll1$ since $a_{\tpb_{\rm min}}\ll 1/\tpsi$).  Thus, since $b_{n}\gtrsim b_{m}$
(with near equality holding once $A\gtrsim1$; see \cref{sub: dbprl derivations}), $b_{y} \hat{v}_{{\rm A}y} $
dominates for $a\gtrsim a_{\tpb_{\rm min}}$, meaning switchbacks
 result from large $b_{y}$ fluctuations, rather than large $b_{x}$ fluctuations, even though $\hat{v}_{{\rm A}x}> \hat{v}_{{\rm A}y}$  for a wide range of $a$ after this point (until $a\lesssim 1/\tpsi$). 
 This behavior can be seen in the second ($a=10$) panel of \cref{fig: ps solutions}, which 
is pictured slightly before $a_{\tpb_{\rm min}}\approx 12.7$ for these parameters: at this $a$, some $b_{\|}$ minima are dominated
by $b_{y}$ fluctuations (e.g., around $\lambda\approx 0.1$), some are dominated by $b_{x}$ fluctuations (e.g., around $\lambda\approx 0.45$), while the largest $b_{\|}$ perturbations involve both (e.g., around $\lambda\approx 0.7$).
In contrast, by later times (e.g., the next panel where $\tps \approx 30^{\circ}$), $b_{\|}$ fluctuations are
nearly completely determined by large changes in $b_{y}$ to $b_{y}<0$.

 With this piece of information in hand --- that the enhanced switchback formation from the
 Parker spiral involves switchbacks that are dominated by  $b_{y}$ perturbations --- 
 it is straightforward to demonstrate that $b_{x}$ must increase through a switchback. 
First,
 we form the spatial constant 
 \begin{equation}
\Delta b^{2} = (\ba+\va)^{2}-\va^{2} = 2v_{{\rm A}x}b_{x} + 2v_{{\rm A}y}b_{y} + b_{x}^{2}\csc^{2}\tp + b_{y}^{2},
\end{equation}
which is the difference between the total and mean-field magnitude (we use \cref{eq: div b for z sb} for $b_{z}$). 
Solving for $b_{x}$ gives
\begin{equation}
\frac{b_{x}}{\sin^{2}\tp} = -v_{{\rm A}x} + \sqrt{v_{{\rm A}x}^{2} + \csc^{2}\tp\left[ \Delta b^{2}+v_{{\rm A}y}^{2} - (b_{y}+v_{{\rm A}y})^{2}\right]}.\label{eq: bx soln for one sided}
\end{equation}
Here,  $\Delta b^{2}+v_{{\rm A}y}^{2}$ is simply a positive constant, while $b_{y}+v_{{\rm A}y}$ is the total $y$-directed field.
The key insight from \cref{eq: bx soln for one sided} is  that $b_{x}$ is a monotonic function of $(b_{y}+v_{{\rm A}y})^{2}$ and is maximized when $(b_{y}+v_{{\rm A}y})^{2}=0$. 
Then, recall that $b_{\|}$ changes are driven  by $b_{y}\hat{v}_{{\rm A}y}$, while $\hat{v}_{{\rm A}y} $ is rather small,
implying that any large change to $b_{\|}/\vam$ must involve $b_{y}$, and thus $b_{y}+v_{{\rm A}y}$, passing through zero (since $b_{y}$ must also clearly remain less than the total field magnitude). 
Thus, any large change to $b_{\|}$ must occur around the same location as $b_{x}$ being maximized,
which implies that the magnetic-field vector rotates through the positive-radial direction during the switchback
(note that $b_{x}$ can subsequently decrease as $b_{y}$ becomes large and negative and $(b_{y}+v_{{\rm A}y})^{2}$ increases).
This feature is highlighted by the grey shading in \cref{fig: ps solutions}, which show the regions where $|d b_{\|}/d\lambda|$ is near its maximum in each panel; clearly, such regions  generically line up with positive peaks in $b_{x}$. Physically, all that \cref{eq: bx soln for one sided} is saying is that the constant-$B$ constraint implies that if $b_{y}+v_{{\rm A}y}$ passes through zero, $b_{x}$ must increase to compensate, even though 
if it decreased instead it could in principle help to form switchbacks. It is also worth mentioning that \cref{eq: bx soln for one sided} remains valid even for radial $\va$ or for $a\lesssim 1/\tppi$, when $b_{\|}$ is instead dominated by the contribution from $b_{x}\hat{v}_{{\rm A}x}$; but, in this regime it does not provide any obviously useful constraint.

Also of interest is that \cref{eq: bx soln for one sided} excludes the possibility that the field rotates beyond $90^{\circ}$
in the $+\hat{\bm{y}}$ direction (i.e. the +T direction) at all. Such a rotation would involve 
$(b_{y}+v_{{\rm A}y})^{2}$ reaching a maximum value then decreasing, while $b_{x}$ would have to continuously decrease, which is impossible due to the monotonic dependence of $b_{x}$ on $(b_{y}+v_{{\rm A}y})^{2}$. Put
together with the discussion of the previous paragraph, this provides an alternate way to consider the 
switchback skewness: a field rotation from $\vah$ in the +T ($+\hat{\bm{y}}$ direction) is limited to be less than $90^{\circ}-\tps$ when projected onto the $x,y$ plane; but, a field rotation  in the -T direction can rotate the field by $90^{\circ}+\tps$. This leads to a strongly asymmetrical distribution of field rotations. This feature is clearly seen in the 3-D simulations of  \zades\  (see their figure 7-9), despite the fields therein obviously not satisfying the 1-D approximation used to 
derive \cref{eq: bx soln for one sided}.

\subsection{Switchback sharpness, irregularity, and the direction of the mean field}\label{sub: sb intermittency}

\begin{figure}
\begin{center}
\includegraphics[width=1.0\columnwidth]{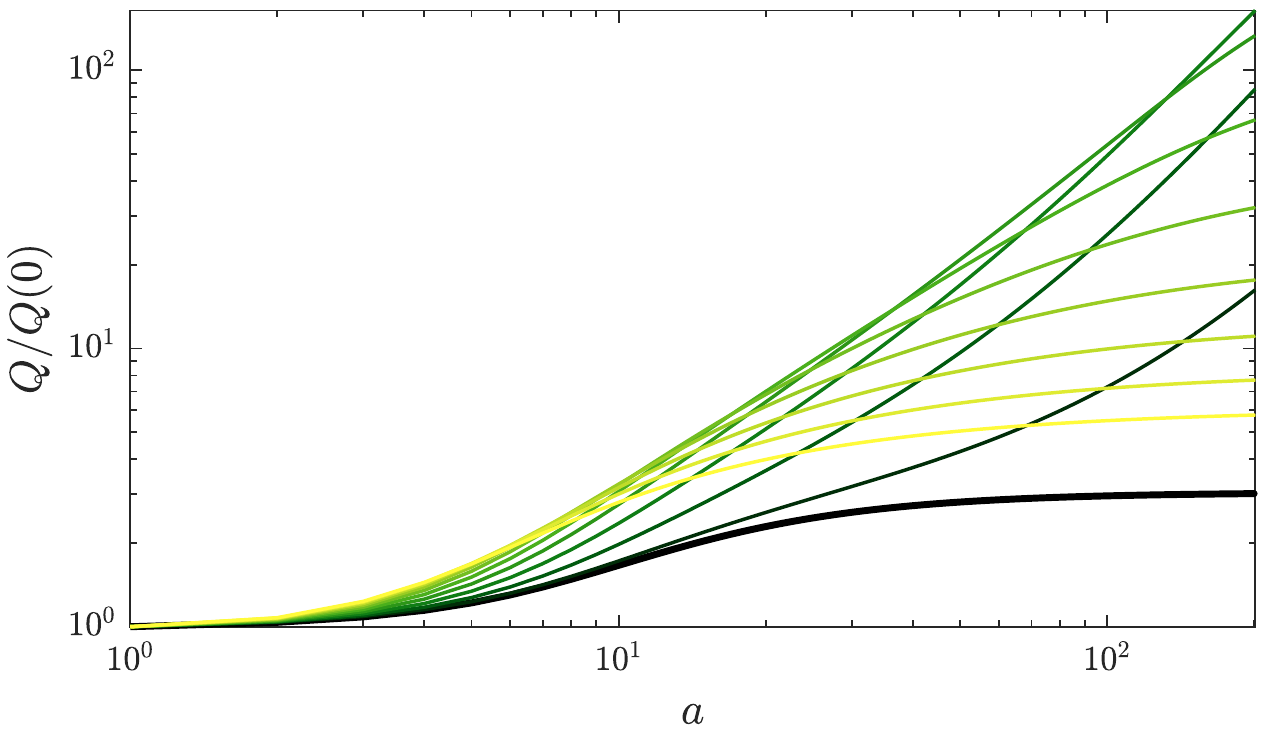}
\caption{Waveform sharpness, as measured by ``steepening factor'' $Q\equiv \overline{|\partial_{\lambda}\ba|^{2}}/\overline{|\ba|^{2}}$ for the
same set of waves as shown in \cref{fig: ps sbs from waves} (evolved using \cref{eq: alfreds b eqn,eq: alfreds u1 eqn}). The color 
scheme is the same, with the thick black  line showing the radial-$\va$ case and the colors showing increasing $\tpsi$ in steps of
$2^{\circ}$ from dark green to yellow. In a radial mean field, there is very little generation of sharp structures, as is clear from \cref{fig: no ps solutions}. 
The Parker spiral significantly changes this, causing the development of much sharper waves even after relatively short evolution times (see \cref{fig: ps solutions}). }
\label{fig: discontinuities}
\end{center}
\end{figure}

The most obvious feature of the solutions shown in \cref{fig: ps solutions} is how sharp and irregular
the switchbacks become, \emph{viz.,} solutions feature wide quiet sections interpersed by sudden and rapid 
switchbacks as $b_{y}$ changes sign. That these solutions evolve to become significantly sharper and more sporadic than those 
with radial $\va$ is clear from a quick by-eye comparison with \cref{fig: no ps solutions}, which shows 
wave evolution in a radial mean field with the same initial $b_{n}$. We 
also show in \cref{fig: discontinuities} the ``steepening factor'' $Q\equiv \overline{|\partial_{\lambda}\ba|^{2}}/\overline{|\ba|^{2}}$ (\alfreds), for the same set of 1-D waves as shown in \cref{fig: ps sbs from waves},
which demonstrates the same idea more quantitatively and shows how the effect is even stronger for larger $\tps$.
 Promisingly, we see  evidence that
this effect of the Parker spiral enhancing switchback sharpness carries over to fully 3-D solutions (see  figure 5 of \zades), which suggests it may be observable in the solar wind. 

To understand why the effect occurs, we must simply couple the fact that $\overline{b_{y}}=0$ 
to the conclusion of the previous paragraph that $\pm$T field rotations are limited to 
 angles $< 90^{\circ}\mp\tps$ from the mean field (projected onto the radial-tangential plane).
As the Parker spiral rotates further, the  difference between these two directions increases further, 
but $\overline{b_{y}}$ remains zero, implying that the field must spend more volume rotated in 
the +T direction than the -T direction to compensate for the limited size of the +T rotations. 
The consequence is clearly seen in the bottom half of \cref{fig: ps solutions} (for $a\gtrsim 15$ or so, once $\tps$ becomes significant): the solutions spend wide regions with modest $b_{y}>0$, then suddenly rotate to $b_{y}<0$ over short distances. 
This leads to both sharper waveforms and a more irregular, intermittent distribution of $b_{\|}$ fluctuations.

A  corollary of the previous paragraph's discussion relates to measuring the Parker spiral direction. In particular, 
the mode of the field-direction  (its most-common direction) becomes quite different to the mean-field direction $\vah$, which is the 
direction in which the waves actually propagate. This is particularly clear from e.g., the $a=30$ panel in \cref{fig: ps solutions}, which has $\tps\approx 45^{\circ}$, but, discounting the switchbacks, one would conclude $3v_{{\rm A}x}\approx v_{{\rm A}y}$, or an angle of $\approx 70^{\circ}$. This feature is also clearly seen in the simulations of \zades\ (see their figure 7),
suggesting that in measuring the Parker spiral angle in the solar wind, one must be careful to resolve the distinction between 
statistical mean and mode. 

Finally, we note that the comparison between \cref{fig: ps sbs from waves,fig: ps solutions,fig: discontinuities} reveals an interesting consequence of the intermittent switchback rotations. Once $\tps$ rotates beyond $45^{\circ}$ and $A$ starts decreasing, 
 fluctuations in $b_{\|}/\vam$ also decrease, if measured  by their root mean square deviation (see \cref{fig: ps sbs from waves}, which matches the prediction \cref{eq: dbprl}). However, we see from \cref{fig: ps solutions} that 
 this decrease does not involve individual switchbacks becoming smaller, but
 rather a decrease in their volume-filling fraction. This is clear from the fact that the relative size of individual $b_{\|}/\vam$ fluctuations increases from $a=50$ to $a=200$ in \cref{fig: ps solutions}, and that $Q$ (the waveform steepness) continues
 to grow rapidly at large $a$ (\cref{fig: discontinuities}). This 
 trend appears to continue up until arbitrarily large $a$ for waves with $\tppi\gg\tpsi$, with nominally 
 small-$A$ solutions containing extreme and sudden changes in field direction. Nonetheless, although interesting, 
 we do not expect this property to be particularly important to the solar wind: by such large $a$ with $\tps\gtrsim 45^{\circ}$, 
there has likely been significant reflection of forward into backwards propagating waves and turbulence, which 
in general seems to destroy the constant-$B$ requirement needed to form such solutions (see \cref{sub: turbulence discussion} below and figure 3b of \zades).

\section{Discussion: heuristic application to 3-D fields}\label{sec: discussion}

In this section, we provide some commentary on how our results can be 
applied to fully 3-D fields, as needed for application to the solar wind. 
We also, in App.~\ref{sub: integral estimate}, provide a different --- more generic but less informative --- argument 
for the results from \cref{sub: sb growth in ps} that switchbacks are enhanced by the Parker spiral and form preferentially 
with  field rotations  that lie in its plane ($\ph^{\rm (Z)}$ in the out-of-plane direction). 
A key idea in this argument, as well as for the qualitative discussion, is that
expansion generically tends to expand structures in the perpendicular plane (i.e., make them pancake shaped), 
or equivalently, to rotate the wave vectors to become more radial. In the presence of turbulence --- 
which, to the contrary, tends to elongate structures along the background magnetic field --- the competition 
with expansion will presumably enhance the power in radially aligned (as opposed to field perpendicular) wavevectors, compared to turbulence
without expansion. Indeed, this seems to be observed in EBM simulations \cite{Verdini2015}, and likely also in the solar wind \cite{Chen2012,Verdini2019}.

In order to appraise the application to 3-D fields  in more detail, let us start 
by pointing out that there are (at least) three important differences in 3-D that are lacking from the 1-D results
above. The 
first and most obvious is structure: simply the fact that fields vary in all three directions, not just in $\p$. 
The second is reflection-driven turbulence, which cannot affect 1-D fields  because the nonlinear 
term vanishes (indeed it is this feature that enables the derivation of \cref{eq: alfreds b eqn}; \alfreds), 
but will in general cause the destruction of the pure Alfv\'enic state, the decay of fields compared to the
WKB expectations, and the (re-)population of power across a wide range of wavenumbers. 
The third is parametric decay \cite{Sagdeev1969,Cohen1974a,Goldstein1978} --- i.e.,  instability of the nonlinear Alfv\'enic solutions --- which 
can destroy the wave if its perturbations grow sufficiently large. While this can  occur in 
1-D also, it is convenient to discuss it here  because it is not captured by our analysis or by \cref{eq: alfreds b eqn,eq: alfreds u1 eqn}.  Let us address each of these in turn. 

\subsection{Structure} 

Realistic solar-wind fields presumably involve power across a wide range of wavenumber directions, distributed in such a way as to ensure constant $B^{2}$. 
We suggest that because the key physical ingredients needed for most of our results above are relatively simple --- $\nabla\cdot\ba=0$, $B^{2}={\rm const.}$, and 
the driving of $\p$ towards the radial direction --- these results can also apply in  
3-D with important caveats. In this application,  the $\p$ direction should correspond to the direction of fastest variation across some particular substructure
of the 3-D field. 
 The simplest example of this viewpoint  is from \cref{eq: dbprl}, which shows that near-perpendicular
wavenumbers generate large $b_{\|}$ perturbations because of the the constant-$B$ constraint. As discussed in \alfreds, the application 
to 3-D fields is straightforward, implying that structures that vary rapidly in the nearly perpendicular direction --- in other words, those that are elongated along the mean magnetic field --- drive larger $b_{\|}$. This feature
has been seen in both simulations \cite{Shoda2019} and observations \cite{Laker2021}.

A similar application is to the conclusions of \cref{sub: sb growth in ps}, that a Parker spiral should enhance
switchback formation, and that switchback field deflections occur preferentially
in the tangential plane. As demonstrated in more detail  in App.~\ref{sub: integral estimate}, when
structures are compressed to perpendicular pancakes by expansion ($p_{x}\gtrsim p_y,\,p_{z}$ on average), wavenumbers are on average more perpendicular 
to a mean field with a  Parker spiral than a radial mean field, causing enhanced switchback occurrence rates by \cref{eq: dbprl}. Similarly, if   $\va$ involves a Parker spiral 
and the wavenumber distribution is biased towards the radial direction, 
wavenumbers that
lie perpendicular to the plane of Parker spiral ($\ph^{\rm (Z)}$ waves) are on average more  perpendicular to $\va$ 
than those in the plane of the Parker spiral. Even in 3-D, Alfv\'enically polarized fluctuations involve larger field 
perturbations perpendicular to their direction of fastest variation, implying that the structures that generate larger switchbacks are more likely to involve tangential rotations of $\ba$ in the plane of the Parker spiral (i.e., $b_{y}$).
Indeed, both of these features are seen in the simulations of \zades, with their figure 6 showing the stronger switchback
growth with a Parker spiral, and their figures 7-9 showing various measures of deflections becoming tangentially asymmetric. 
Tangential asymmetry of deflections is also seen quite clearly in PSP data \cite{Laker2022}.  Nonetheless, turbulence provides an important caveat, particularly to the conclusion about switchback growth (see below).

More complex are the conclusions of \cref{sec: structure}, where we showed that waves become strongly tangentially skewed. 
As well as $\nabla\cdot\ba=0$ and constant $B^{2}$, this conclusion relies on the idea that the $b_{\|}$ of a   switchback becomes dominated by the $b_{y}$ contribution, rather than the radial ($b_{x}$) fluctuation. This in 
turn relied on the wavevector becoming predominantly radial as $\va$ rotates away from the radial (see  \cref{eq: bprl at amin}). If these conditions are satisfied,  \cref{eq: bx soln for one sided} suggests that structures that
vary fastest in the near-radial direction  ($\tp\lesssim 45^{\circ}$), but nonetheless remain somewhat perpendicular to $\va$ ($\tpb\gtrsim 45^{\circ}$),  must increase their radial-field perturbation through a switchback in order to maintain constant $B$. This  makes  them tangentially skewed, therefore causing a significant difference between the statistical mean and mode 
of the magnetic-field direction. This  condition --- that quasi-radial $\tp\lesssim 45^{\circ}$ structures start dominating switchbacks for modest $\tps\lesssim 45^{\circ}$ --- does not seem unreasonable, so long as the turbulence is 
not continually repopulating modes along $\va$ as fast as they are being rotated radially by expansion (see below). 
Indeed, such tangential skewness is undeniably obvious in the Parker spiral simulation of \zades\ (see their figures 7c and 8), which provides at least a basic confirmation of the above scenario.
The feature is less clear in a recent analysis of PSP deflections \cite{Laker2022}, although it seems to be present in some cases (most prominently, encounter E6) and will be influenced by the Parker spiral angle,  the fluctuation's amplitude, and the analysis method (e.g., the deflection angles that are counted as switchbacks).  
The enhanced switchback sharpness  (\cref{sub: sb intermittency}) seems to result from $\sin\tpb$ being an 
increasing function of time, so presumably applies under similar circumstances with similar caveats. This feature 
is also observed in figure 5 of \zades, with the Parker spiral simulation exhibiting more sharp field rotations.

\subsection{Turbulence}\label{sub: turbulence discussion}

Turbulence in the solar wind is thought to be caused by (among other possibilities) the reflection of outwards- to inwards-propagating perturbations  \cite{Velli1989}, whose amplitudes we will term $z^{+}$ and $z^{-}$, respectively. If this happens sufficiently 
rapidly so as to to cause the ratio $z^{-}/z^{+}$ (the imbalance)  to grow continuously,
the process will destroy the nonlinear solution \cref{eq: nl aw solution}, eventually breaking the constant-$B$ condition
and invalidating all of our arguments above. Indeed, the start of this process can be observed at late times in most simulations of \zades, with $C_{\bm{B}^{2}}$ (a measure of the relative spherical polarization) increasing at late times as the imbalance decreases (see their
figures 2b \& 3b). In the solar wind, such a process is at least only partially complete 
by ${\sim}1{\rm AU}$, where turbulence is still observed to be relatively imbalanced and spherically polarized \cite{Bruno2013}; nonetheless, all of our results for $\tps\gtrsim 45^{\circ}$ (e.g., the bottom two panels 
of \cref{fig: ps solutions}) are clearly suspect and will likely be invalidated by this effect. However, even well before $z^{-}\sim z^{+}$,  turbulence causes two other effects that invalidate our arguments if they are sufficiently strong: the first is the turbulent decay of the magnetic field; the second is the re-population of
wavenumbers through nonlinear interactions. Turbulent decay will decrease the growth of wave amplitude below  $A\propto a^{1/2}$, as used in our estimates (for $\tps\ll45^{\circ}$); clearly if waves stop growing there will likewise 
be no growth of switchbacks (unless perhaps if $A\gtrsim1$ already and $\sin\tpb$ increases). Similarly, if the interaction between different modes $\p$ is stronger than  the effect of expansion, 
the scaling of $\p$ used in the arguments above will be incorrect (although there will presumably be some expansion-driven bias 
towards the radial direction). While this does not necessarily hinder switchback formation --- indeed, stopping the decrease in $\sin\tpb$ would be helpful --- it would at least invalidate our scalings. 

\zades\  argued (see their \S II \!C), based on previous work \cite{Verdini2007,Chandran2009,Chandran2019}, that the importance of the effects discussed above should be determined by the parameter $\chi \approx k_{\perp} z^{+}/k_{\|}\vam \approx A k_{\perp}/k_{\|}$, which is a measure of the relative size of nonlinear effects ($k_{\perp}z^{+}$) and wave propagation ($k_{\|}\vam$) for the $z^{-}$ waves 
(here $k_{\perp}$ and $k_{\|}$ should be interpreted as  average inverse correlation lengths of the energetically 
dominant $z^{+}$ structures, which source $z^{-}$ through reflection). For $\chi\gtrsim1$, the phenomenology suggests turbulent decay balances 
expansion-induced growth such that $A\propto a^{0}$ and waves do not grow at all (this is tentatively supported by the results of  a $\chi>1$ simulation in \zades; see their figure 2a). In this 
case, our results likely do not apply. In contrast, for $\chi\lesssim1$, 
turbulent effects are weaker, and many of our results for 1-D waves seem to apply relatively well to 
3-D, as evidenced by the multiple items of agreement discussed above. This conclusion --- that 
the applicability  our results in 3-D is determined by $\chi$ --- 
requires further study and is currently quite poorly understood. For instance,  the phenomenology seems to predict faster decay than seen in simulations and observations \cite{Chandran2009,vanBallegooijen2016,Chandran2019}, and kinetic effects could also play a dominant 
role if they halt turbulent decay \cite{Meyrand2021}.

Finally, it is worth reiterating that these conclusions about turbulence, and indeed all of our results, apply
only to regions of super-Alfv\'enic wind beyond the Alfv\'en surface because they use the expanding box model. In the
sub-Alfv\'enic wind, fluctuation amplitudes almost certainly grow robustly even in the presence of turbulence \cite{vanBallegooijen2011,Perez2013}, while
without turbulence waves grow much more rapidly than \cref{eq: A evolution with PS}. In addition, if we interpret 
$a$ as the cross-sectional area of a flux tube (see \zades\ \S II \!D), the 
scaling of $\p$ and $\tps$ with $a$ is quite different in sub-Alfv\'enic regions \cite{Weber1967,Tenerani2017}, and of course $\tps\ll1$ in such regions anyway. Thus, for all our results, it is imagined that fluctuations have arrived 
already with a relatively large amplitude, and perhaps switchbacks, at the Alfv\'en point. It does not seem
possible to grow large switchbacks from very small amplitude fluctuations in the EBM because of the turbulent decay
(see \S II \!C of \zades). 

\subsection{Parametric decay}

A third physical aspect that is missed out by our analysis is parametric decay, \emph{viz.,} instability 
of the nonlinear Alfv\'enic solution \eqref{eq: nl aw solution} \cite{Goldstein1978}. This can afflict even 1-D waves, however it is  not captured by the reduced equations  \eqref{eq: alfreds b eqn}--\eqref{eq: alfreds u1 eqn} on which we base our analysis. In general, it can cause break up of the wave if the instability grows to saturate at large amplitudes.  Its
 growth rate, which will determine the time before 
 saturation,  generally increases with wave amplitude and at lower plasma $\beta$. For both general parallel fluctuations \cite{Malara2000} and oblique large-amplitude 
 waves of the form discussed in \cref{sub: forming constant B solns} \cite{DelZanna2001},
 parametric instability has been found to be quite virulent, leading to saturation with $z^{+}\sim z^{-}$. 
 However, Ref.~\onlinecite{DelZanna2015} found that expansion strongly decreased the growth rate of the instability in the
 EBM, which is likely due to the dynamics slowing down as the plasma expands, suggesting waves with frequencies $\omega_{\rm A}$ not too far above $\dot{a}/a$ could propagate undisrupted over a relatively wide range of $a$
  (in contrast Ref.~\onlinecite{Tenerani2013} find
parallel waves are rapidly destroyed in the inner heliosphere, so more work is needed to better understand the influence of expansion).
 Perhaps more importantly, there are hints that the instability saturates at much lower levels in 2-D or 3-D fields  
 \cite{Primavera2019,Tenerani2020},  maintaining the nonlinear Alfv\'enic state \eqref{eq: nl aw solution} nearly unchanged with $z^{-}\ll z^{+}$ even after saturation  (although the instability may still play a key role in seeding turbulence \cite{Shoda2019}).
 In addition, parametric decay is at least partially stabilized by damping of compressive fluctuations (which should be strong in a collisionless plasma), and random
  structure in the background Alfv\'enic state \cite{Cohen1974a}.
 Overall, more study is needed to better understand the relevance of parametric instability compared to reflection-driven turbulence, 
 but if it either grows too slowly or saturates at low levels, it will not significantly change our results. 
 
\section{Conclusions}\label{sec: conclusions}

In this paper, we have explored the influence of the Parker spiral on 
the evolution of Alfv\'enic switchbacks in an expanding plasma.
Using simple, geometric arguments based on spherically polarized 1-D waves, we find a surprisingly large effect. 
This highlights the interesting, nonintuitive physics of nonlinear Alfv\'enic perturbations, underscoring
that particular care must be taken before attributing any observed asymmetrical (or otherwise unexpected) characteristics of switchbacks to properties of their source. 
The key differences compared to the case with a radial mean field all result from the nontrivial (non-monotonic) evolution 
of the wave's obliquity $\sin\tpb$, which is brought by the simultaneous 
rotation of the mean field $\va$ and mode wavevector $\p$ in different directions. 
Surprisingly, despite the normalised amplitude of waves in a Parker
spiral growing more slowly than with a radial mean field, the formation of 
switchbacks is strongly enhanced in the most relevant regimes (c.f.~black and green lines in \cref{fig: theory ps sbs}). This conclusion
may be testable in the solar wind by comparing streams with different mean-field directions but similar 
fluctuation amplitudes. Our other main conclusions are as follows:\vspace{0.1cm}\\
(i) Switchbacks preferentially involve field rotations in the tangential direction, \emph{viz.,} a rotation 
in the plane of the Parker spiral. This is because wavevectors perpendicular to the plane of the Parker
spiral ($\ph^{\rm (Z)}$ or $\cos\varphi=0$) are more effective at generating parallel field perturbations
than those in the plane of the Parker spiral  ($\ph^{\rm (Y)}$ or $\cos\varphi=\pm1$). This conclusion
is based on both the evolution of 1-D waves (\cref{sub: sb growth in ps}) and a simple argument based on
the average obliquity of wavevectors in a spectrum biased by radial expansion (\cref{sub: integral estimate}).\vspace{0.1cm}\\
(ii) Tangential switchbacks (those discussed in conclusion (i)) become strongly ``tangentially skewed,''
meaning the sharp field rotations of the switchback preferentially occur in one direction (towards the radial 
component of the mean field). This is a consequence of the divergence-free constraint in a spherically polarized wave, 
which forces the radial-field fluctuation to increase, rather than decrease, as the tangential field passes through zero (\cref{sub: proof of skewness}). 
A similar constraint is that, projected on to the radial-tangential plane, field rotations in the $\pm$T direction are limited to
angles $\leq 90^{\circ}\mp \tps$, where $\tps$ is the Parker-spiral angle; thus $-$T field rotations can be significantly larger,
causing a highly asymmetrical rotation distribution.\vspace{0.1cm} \\
(iii) As a consequence of conclusion (ii), and in order to maintain mean-zero fluctuations, the 
field-direction mode (its most common direction) is strongly skewed towards the $+$T direction compared to
the mean field, \emph{viz.,} it usually rotated to a larger angle than $\tps$. This suggests that care must be taken 
in measuring the Parker spiral, which should be the mean-field direction (that being the direction in which Alfv\'enic perturbations propagate). \vspace{0.1cm}\\
(iv) As another consequence of conclusions (ii) and (iii), switchback fluctuations in a Parker spiral become 
more intermittent and sharper than those in a radial mean field. There are long quiet periods in which 
the field is rotated beyond $\tps$, interspersed with short and sudden large rotations  (see \cref{fig: ps solutions}).\vspace{0.1cm}

Although these conclusions are clearly limited by our reliance on 1-D wave 
physics, we provide an extended commentary in \cref{sec: discussion} about the general 
applicability to 3-D fields with turbulence. This suggests that there can be significant caveats, sometimes to the point of
nullifying most of our results (e.g., in regimes where turbulence nonlinearities completely dominate over expansion), 
but also regimes where we might expect our results to  apply qualitatively, even in 3-D. More importantly, 
in our companion paper \zades, we see evidence for all five of the above conclusions (enhanced switchbacks in 
the Parker spiral, plus each of points (i)--(iv) above) in full 3-D compressible expanding-box MHD simulations. 
Conclusions (ii)--(iii), on the skewness of tangential switchbacks, are particularly clear in field-deflection 
distributions (see figures 7-9 of \zades). There may also be tentative evidence for observations
of these features in PSP and other spacecraft data: Refs.~\onlinecite{Horbury2020,Laker2022} see enhanced numbers of switchbacks 
with tangential deflections per point (i) above; Ref.~\onlinecite{Schwadron2021} reports that switchbacks preferentially deflect 
to one side as per point (ii) above (the feature is less clear, though plausibly present, in the analysis of Ref.~\onlinecite{Laker2022}). The other predictions are also potentially observable --- for instance, 
one could compare switchbacks between streams with different Parker spiral angles to investigate point (iv) or the
overall prevalence of switchbacks --- but require further work. 

\acknowledgements
The authors thank R.~Laker, T.~Horbury, and S.~Bale for interesting discussions in the course of this work.
Support for  J.S. was
provided by Rutherford Discovery Fellowship RDF-U001804, which is managed through the Royal Society Te Ap\=arangi. 
Z.J. was supported by a postgraduate scholarship publishing bursary from the University of Otago.
A.M. acknowledges the support of  NASA through  grant 80NSSC21K0462.
R.M. was supported by Marsden fund grant MFP-U0020, managed through the Royal Society Te Ap\=arangi. 

\appendix

\section{Preferential tangential deflections in 3-D fields due to expansion}\label{sub: integral estimate}

In this appendix, we provide an alternate argument for two of the main 
results of \cref{sec: SB formation due to expansion}: (i) that the Parker spiral tends to enhance switchback 
formation, and (ii) that switchbacks in a Parker spiral tend to involve tangential, rather than normal, deflections. 
Our method is simply to posit a simple Gaussian form for a spectrum of waves with its axis aligned 
along the radial direction, then compute the average $\sin^{2}\tpb$ formed by such a spectrum 
when the mean field lies at angle $\tps$ to the radial. 
In reality, of course, the spectrum should 
be neither Gaussian nor aligned perfectly along the radial: its alignment will presumably result 
from a competition between the expansion, which would tend to create spectral contours that
are elongated along the radial direction, and turbulence, which tends to create spectral contours that 
are pancake shaped about the mean field. Nonetheless, the argument is relatively simple, does not have 
strong dependence on the chosen functional form of the spectrum, and can be trivially extended 
to account for nonradial alignment of the spectrum by simply redefining $\tps$ as the angle between
$\va$ and the spectrum's axis of symmetry (if this exists).

\begin{figure}
\centering
\includegraphics[width=1.0\columnwidth]{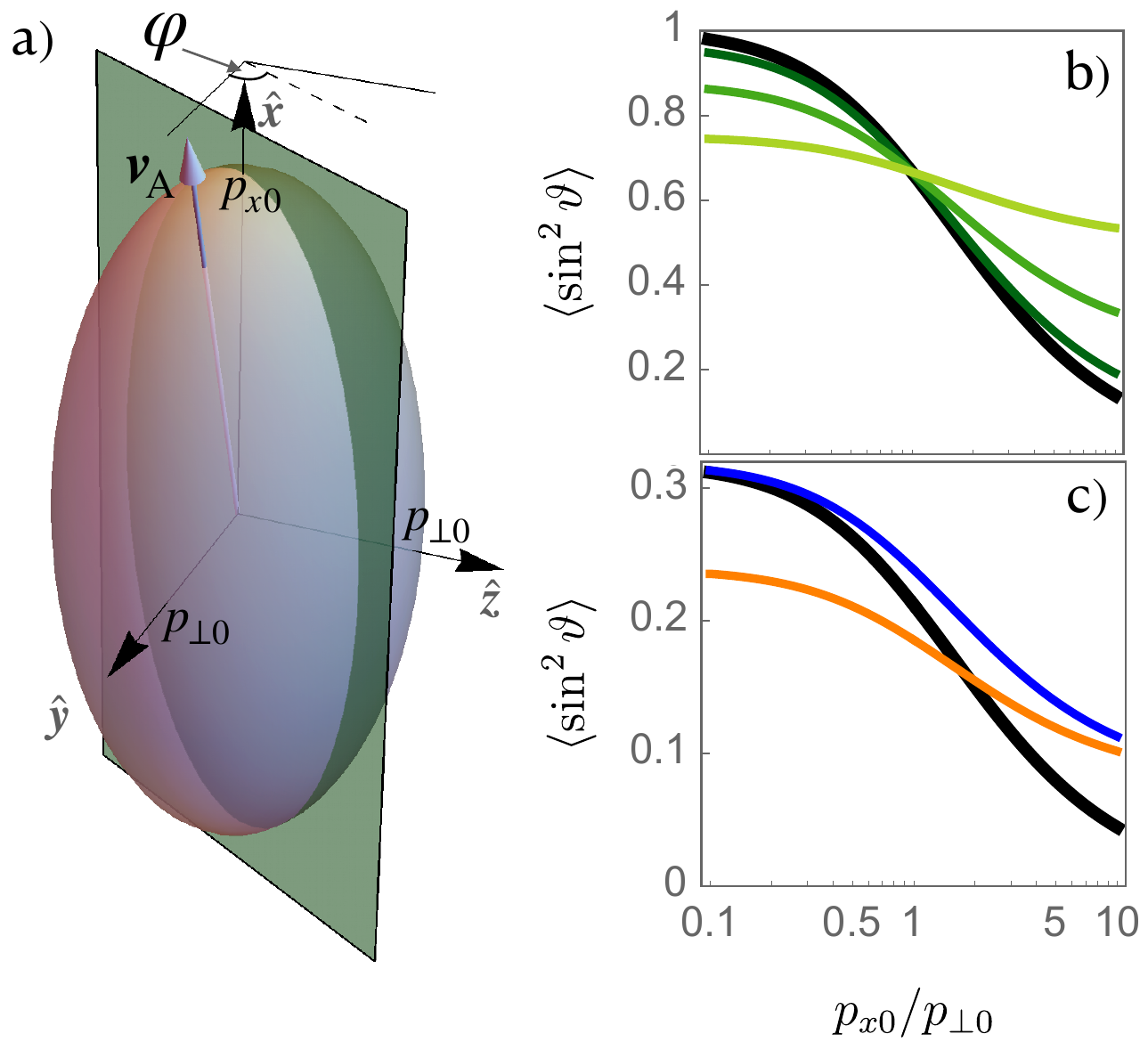}
\caption{Estimates of the switchback prevalence occurring due to a Gaussian spectrum of waves that is
misaligned with the $\vah$ direction. Panel a) shows the geometry, with a representative contour of the
spectrum \cref{eq: gaussian spectrum} along with the mean field and the plane slice at angle $\varphi$, which 
is used to estimate the contribution from different deflection directions. Panel b) shows the integral \eqref{eq: integral full}, which 
estimates the average $\sin^{2}\tpb$  of the spectrum
as a function of $\xi_{p} = p_{x0}/p_{\perp0}$ for a radial $\va$ (black curve) and $\tps=15^{\circ},\,30^{\circ},$ and $45^{\circ}$ shown from dark
green to light green. Panel  c) shows the contribution to the integral \eqref{eq: integral full} from fluctuations in the plane of the Parker spiral 
($\varphi=0$ or $\pi$; orange line) or perpendicular to this plane ($\varphi=\pm\pi/2$; blue line), demonstrating 
that tangential field deflections will cause larger switchbacks. The black line shows $\tps=0$ for comparison.}
\label{fig: integral estimate}
\end{figure}

With this idea in mind, the assumed geometry is shown in \cref{fig: integral estimate}a. We take 
the Gaussian  spectrum, 
\begin{equation}
E(p_{x},p_{\perp}) = \frac{E_{0}}{\pi^{3/2}p_{x0}p_{\perp0}^{2}} \exp\left( -\frac{p_{\perp}^{2}}{p_{\perp0}^{2}} -\frac{p_{x}^{2}}{p_{x0}^{2}}\right),\label{eq: gaussian spectrum}
\end{equation}
where $p_{\perp}^{2} = p_{y}^{2} + p_{z}^{2}$ implies we assume  symmetry about the radial ($x$) axis.
As in the main text, we take $\va = \vam (\cos\tps,\sin\tps,0)$, where $\tps$ is the Parker spiral angle.
The spectral properties are then  specified by the anisotropy $\xi_{p}\equiv p_{x0}/p_{\perp0}$, with 
$\xi_{p}>1$ implying that outer-scale eddies  are perpendicularly extended pancakes (elongated in $p_{x}$), and 
$\xi_{p}<1$ implying the opposite \footnote{One can generalize the argument by imagining that the 
competition between turbulence and expansion leads to a spectrum
with its axis aligned between the radial and $\vah$ directions. Then, our calculation follows in the 
same way with $\tps$ redefined as the angle between the symmetry axis of the spectrum and the Parker spiral. 
However, the assumption  of isotropy in $p_{\perp}$ is more questionable in this case, and one should 
really allow for differing $p_{z}$ and $p_{y}$ widths. Nonetheless, given we lack a good model of 
how this competition between turbulence and expansion manifests, while numerical integrations suggest that this does not make a significant
difference to our qualitative conclusions anyway, it does not seem worthy of detailed exploration.   }. \Cref{eq: dbprl} says that for large-amplitude ($A\gtrsim \sin\tpb$) waves,
a measure of the switchback prevalence is $(b_{\|}/\vam)^{2}\sim A^{2}\sin^{2}\tpb$. Equating $E(p_{x},p_{\perp})$ with $A^{2}$ of a given mode at $p_{x},p_{\perp}$, we see  that the relative switchback prevalence for a given amplitude, in the  large amplitude regime of \cref{eq: dbprl}, is $(b_{\|}^{2}/\vam^{2})/A^{2}\sim\langle \sin^{2}\tpb\rangle =E_{0}^{-1}\int d\bm{p} \,E(p_{x},p_{\perp}) \sin^{2}\tpb$. We can evaluate this integral using $\sin^{2}\tpb = 1-(\bm{p}\cdot\vah)^{2}/p^{2}$ by writing $p_{y} = p_{\perp}\cos\varphi$, $p_{z} = p_{\perp}\sin\varphi$ and integrating 
over $p_{\perp}$ and $p_{x}$. This gives
\begin{align}
\langle \sin^{2}\tpb\rangle =  \int_{0}^{2\pi}&  d\varphi \left\{\frac{ \xi_{p}^{2}}{2\pi\tilde{\xi}_{p}^{3}} (\cos^{2}\tps-\cos^{2}\varphi\sin^{2}\tps) \tan^{-1}\!\tilde{\xi}_{p} \right.\nonumber\\
&\left.+  \frac{1}{2\pi\tilde{\xi}_{p}^{2}}\left[\sin^{2}\tps(\tilde{\xi}_{p}^{2}+\cos^{2}\varphi) - \cos^{2}\tps\right]\right\},\label{eq: integral full}
\end{align}
where $\tilde{\xi}_{p}^{2} = \xi_{p}^{2}-1$ and (despite appearances) the expression is valid for both $\xi_{p}>1$ and $\xi_{p}<1$ (the $\tan^{-1}\tilde{\xi}_{p}$ becomes $i\tanh^{-1}|\tilde{\xi}_{p}|$ for $\xi_{p}<1$). 

If we first consider carrying out the $\varphi$ integral in  \cref{eq: integral full}, this allows the 
comparison of the total relative prevalence of switchbacks at different $\tps$. The result is plotted in \cref{fig: integral estimate}b as a function of $\xi_{p} = p_{x0}/p_{\perp0}$. As expected and intuitive from the geometry, for $\xi_{p}>1$ the Parker spiral (green to yellow curves) increases $\langle \sin^{2}\tpb\rangle $, while the opposite occurs for $\xi_{p}<1$. This demonstrates that if eddies become expanded  into perpendicular pancake structures by expansion, the Parker spiral increases the average 
obliquity of the spectrum thus enhancing switchbacks for the same amplitude. This is effectively the same physics
as discussed in \cref{sub: sb growth in ps}, which showed that nonzero $\tps$ increases $b_{\|}/\vam$ significantly for  $a\gtrsim a_{\tpb_{\rm min}}$, which can only occur once $\tp> 45^{\circ}$ in the  relevant regime.

The second conclusion  from \cref{sub: sb growth in ps} was that switchback deflections should be primarily tangential (in $b_{y}$), because many of the wavevectors in the Parker-spiral plane (which create normal field deflections) become highly parallel (\cref{fig: theory ps sbs}b). 
To assess this conclusion for the spectrum \eqref{eq: gaussian spectrum}, we imagine considering the contribution 
of each $\varphi$ in \cref{eq: integral full} separately.  Wavevectors with $\varphi =0$ or $\pi$ will cause larger $b_{z}$ perturbations (because an Alfv\'enic 
perturbation has polarization ${\sim}\ph\times\va$), while those with $\varphi = \pm\pi/2$ will cause large 
$b_{y}$ perturbations. We plot twice the integrand of \cref{eq: integral full} in \cref{fig: integral estimate}c, which captures the contribution 
to $\langle \sin^{2}\tpb\rangle$ from the $p$-space  plane angled at $\varphi$, as illustrated in \cref{fig: integral estimate}a (this normalization is
such that $\pi$ multiplied by the black $\tps=0$ curve in \cref{fig: integral estimate}c yields the same curve in \cref{fig: integral estimate}b). We see that when $\tps = 30^{\circ}$ the 
contribution from $\varphi=0$ or $\pi$ (orange curve) is quite small compared to that from  $\varphi = \pm\pi/2$ (blue curves), for any value of $\chi_{p}$.
This is not surprising, and is indeed rather obvious by inspection of \cref{fig: integral estimate}a; but, it demonstrates mathematically
that in a random collection of waves with a random series of Alfv\'enic deflections, field deflections in the tangential direction (those
with $|\varphi|\approx \pi/2$) will cause larger parallel-field perturbations than deflections in the normal direction.

\section{Comparison to MHD solutions}\label{app: comparison to MHD}

\begin{figure*}
\centering
\includegraphics[width=1\textwidth]{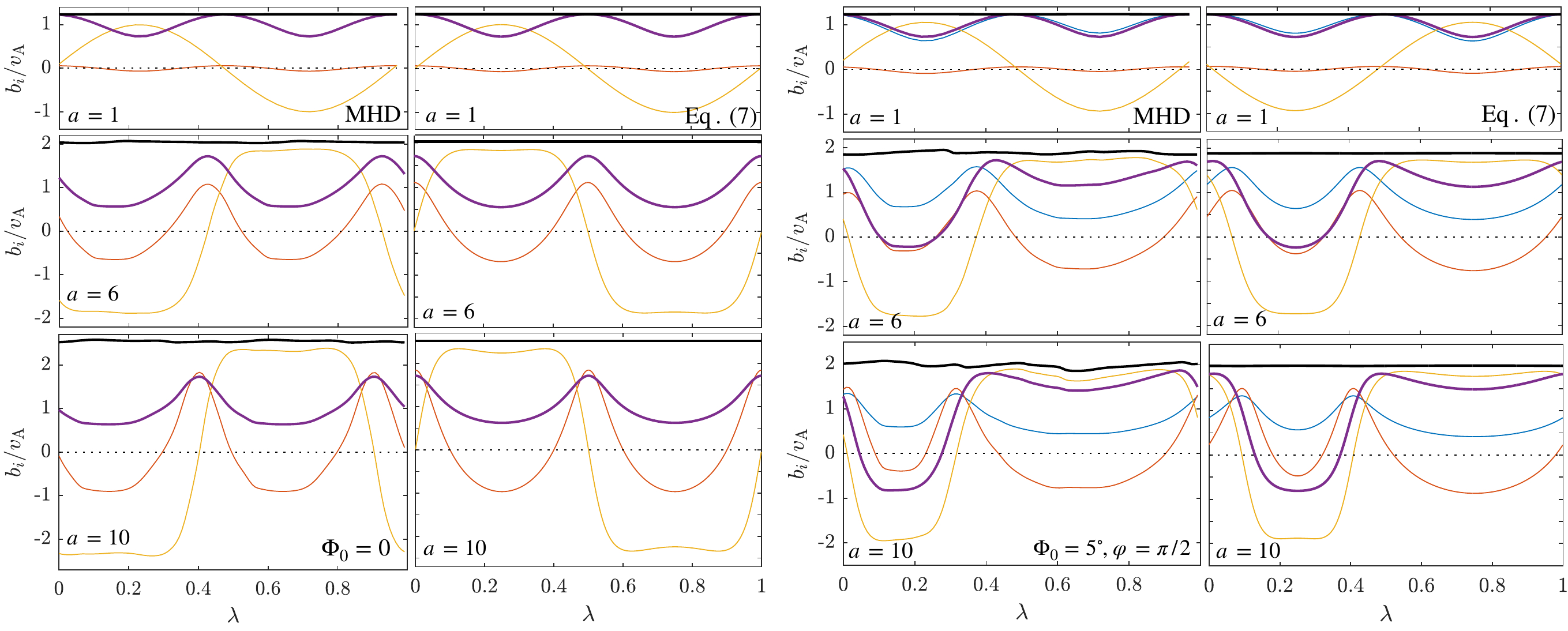}
\caption{Each panel shows the spatial form of an evolving 1-D wave, comparing the wavefront average of true expanding MHD solutions from 2-D simulations (left subpanels; see text) to solutions of \cref{eq: alfreds b eqn,eq: alfreds u1 eqn} (right subpanels). In each case, we start from a sinusoidal $b_{n}$ perturbation, $b_{n}=\sin(2\pi\lambda)$ (giving $A\approx 1/\sqrt{2}$) with $b_{m}$ computed to ensure constant $B$ according to \cref{eq: bm solution for ICs}.
The initial wave obliquity is $\tpi = \tan^{-1}(1/4)\approx 76^{\circ}$, and in the MHD simulations we take $\epsilon = (\dot{a}/a)/\omega_{\rm A} = 0.5$. 
Line colors  in each panel are the same as \cref{fig: ps solutions}: thin blue, red, and yellow show $\hat{v}_{\rm Ax}+b_{x}/\vam$,  $\hat{v}_{\rm Ay}+b_{x}/\vam$, and  $b_{z}/\vam$, respectively, and the thick black and purple lines show the field magnitude $(B/\sqrt{4\pi\rho})/\vam$ and parallel perturbation $b_{\|}/\vam + 1$, respectively. The left-hand set of panels show the case of a radial $\va$, while the right-hand panels include a Parker spiral with $\tpsi=5^{\circ}$ and $\varphi=\pi/2$ (the case  $\ph^{\rm (Z)}$ in \cref{fig: geometry}). We see that the agreement in the evolution, amplitude, and shape is excellent, and that the true MHD solutions
have  reasonably constant $B$ despite the relatively large $\epsilon$ (there are some fluctuations in the MHD wave from the beginnings of parametric instability and compressive perturbations driven by expansion). }
\label{fig: MHD compare radial}
\end{figure*}

In this appendix we directly compare the predictions of \cref{eq: alfreds b eqn,eq: alfreds u1 eqn}, which has been used in the
main text to understand nonlinear wave evolution, to nonlinear isothermal expanding-box MHD simulations with the \texttt{Athena++} code.
The purpose of this comparison is two fold. First, it is simply interesting to better understand the accuracy and applicability of \cref{eq: alfreds b eqn,eq: alfreds u1 eqn}, given 
it was derived through an asymptotic expansion in a slow expansion rate. On this aspect, 
the comparison is extremely positive. Second, we noted in \cref{sec: structure} that 
when $\p$ and $\va$ lie in the same plane (the $\ph^{\rm (Y)}$ case with $\varphi = 0$ or $\pi$), \cref{eq: alfreds b eqn,eq: alfreds u1 eqn} usually
fail to produce solutions with constant $B$. 
Importantly, because \cref{eq: alfreds b eqn,eq: alfreds u1 eqn} are derived by assuming that $\ba$ maintains constant $B$, if it does not, 
we cannot trust their results. It is thus interesting to see whether this production of non-constant $B$ is truly physical --- i.e., whether it also occurs in true MHD evolution --- 
or whether it results for another reason related to the approximations used to derive \cref{eq: alfreds b eqn,eq: alfreds u1 eqn}. 
Although the detailed cause of this behavior remains unclear, we speculate that it relates to  $\sin\tpb$
changing particularly rapidly in this geometry, with  the shape of $\ba$ not able to adjust fast enough to maintain constant $B$.

To generate the MHD solutions, we use the MHD code \texttt{Athena++}, with the  modifications to capture plasma expansion detailed in  \zades.  We set up each wave in a 2-D domain of dimensions $L_{x}=4L_{y}$ at $a=1$, by initialising the sinusoidal 
$p_{x}=2\pi/L_{x}$, $p_{y}=2\pi/L_{y}$ mode in $b_{n}$ (the component out of the plane) with amplitude $1$, such that the wave obliquity is $\tpi = \tan^{-1}(L_{x}/L_{y})\approx 76^{\circ} $. 
The other $\ba$ components are constructed as described in \cref{eq: bm solution for ICs} to ensure constant $B$, with an initial Parker spiral angle of $5^{\circ}$ (if this is included).
We choose the expansion rate to be $\dot{a}/a = 0.5\omega_{\rm A}$ (i.e., $\epsilon=0.5$) and use $128$ grid points in each direction. 
In order to compare the solutions to \cref{eq: alfreds b eqn}, we perform a ``wavefront average'' at each output step, meaning we rotate the coordinate system to align with the wave (accounting for the periodicity of the domain), then spatially average in the direction perpendicular to $\p$.

\begin{figure*}
\centering
\includegraphics[width=1\textwidth]{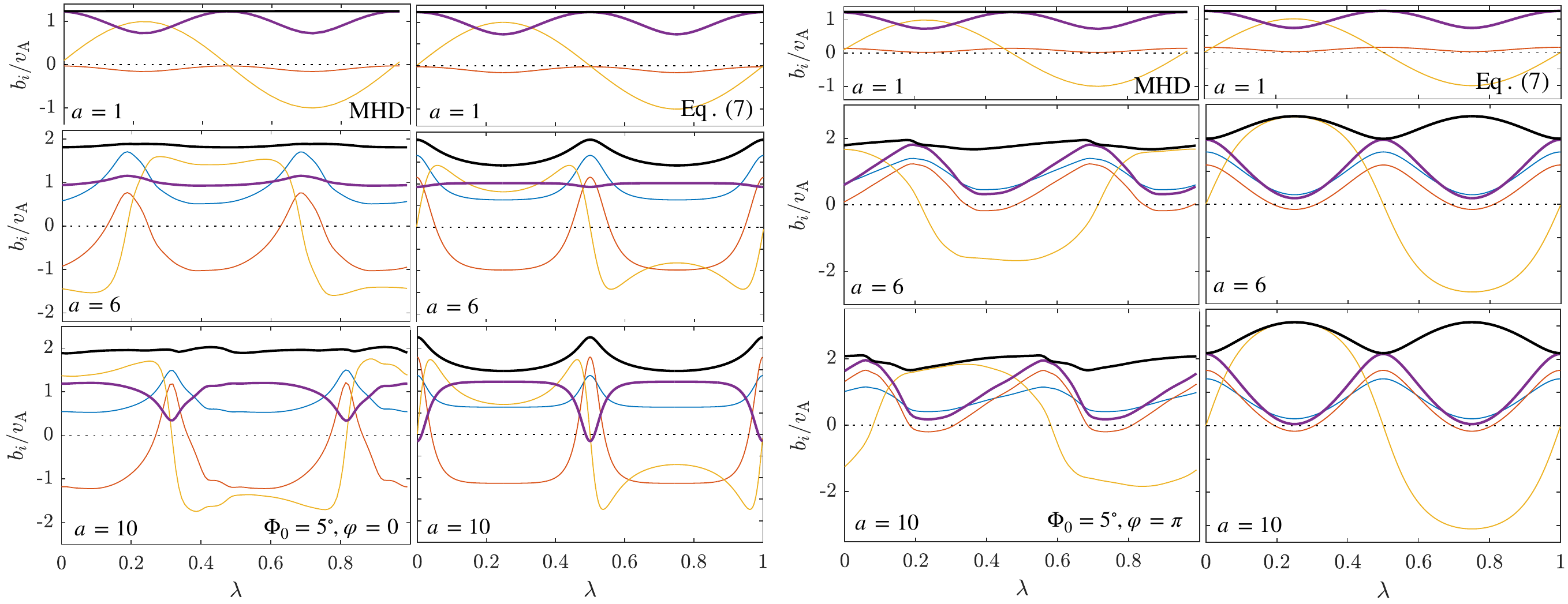}
\caption{As in \cref{fig: MHD compare radial}, with the same initial conditions, $\tpi$, and $\epsilon$, but for the cases $\varphi=0$ (left-hand panels) and $\varphi=\pi$ (right-hand panels), which
both have $\ph$ and $\va$ lying in the same plane (case  $\ph^{\rm (Y)}$ in \cref{fig: geometry}). In these cases, \cref{eq: alfreds b eqn,eq: alfreds u1 eqn} fail to produce constant-$B$ solutions  (see discussion in text), and the difference with the MHD solutions becomes more significant.  
In particular, the MHD solution much more effectively smooths the variation in $B$, generating comparatively smaller switchbacks in both cases.}
\label{fig: MHD compare Y}
\end{figure*}

Results for radial $\va$ and $\varphi=\pi/2$ --- \emph{viz.,} the situations in which \cref{eq: alfreds b eqn,eq: alfreds u1 eqn} 
successfully maintain constant $B$ --- are shown in \cref{fig: MHD compare radial}. 
Left and right subpanels compare wavefront-averaged solutions to the MHD equations to those of \cref{eq: alfreds b eqn} at the same $a$ and other 
parameters.
We see excellent agreement between the general shape of the waveforms, discounting the
phase of the wave, which evolves in the MHD case but not in \cref{eq: alfreds b eqn}. The MHD solutions do involve  small fluctuations 
that are not present in \cref{eq: alfreds b eqn}, which are most clearly observable in the field magnitude profile; these result 
in part from the small compressive components neglected in \cref{eq: alfreds b eqn} (see \alfreds), and in part from 
 the parametric instability (these fluctuations vary across the wavefront direction so are averaged in \cref{fig: MHD compare radial}). The parametric
 instability fluctuations slowly grow and eventually overwhelm the wave, but are not our primary interest here (see Ref.~\onlinecite{DelZanna2015}).

 Results for $\varphi=0 $ and $\varphi=\pi$, when $\p$ and the Parker spiral lie in the same plane, are shown in \cref{fig: MHD compare Y}. 
 In these cases, we can clearly see in the right-hand subpanels that the solutions of \cref{eq: alfreds b eqn,eq: alfreds u1 eqn}
 do not maintain constant $B$, calling the validity of these solutions into question. Indeed, we see  
 that $B$ remains much more spatially constant in the MHD solutions, and the spatial form of the individual components differs
 more significantly than those shown in \cref{fig: MHD compare radial}, although they clearly maintain similar structures. 
Particularly for the $\varphi=\pi$ case, we see tentative evidence that the MHD solutions exhibit  larger variation in $B$ than 
the cases in \cref{fig: MHD compare radial}, suggesting that at least some of the failure of \cref{eq: alfreds b eqn} to maintain constant $B$ is physical. 
In the MHD solutions, these variations propagate  around the box, steepening and reducing in size due to
compressive processes that are not captured by \cref{eq: alfreds b eqn,eq: alfreds u1 eqn} (see, e.g., Ref.~\onlinecite{Cohen1974}). While the situation in the $\varphi=0$ solution is less clear,
 we clearly see smaller switchbacks in MHD in both cases, which presumably results from the combination of MHD more
 robustly maintaining a constant-$B$  Alfv\'enic  solution, and MHD dissipating or smoothing the wave energy that is converted into
 compressive structures. This latter effect would not be captured by the arguments in \cref{sec: SB formation due to expansion}, suggesting
 that the $b_{\|}/\vam$ estimated therein could be an overestimate.
 Also of note is that we do not see any obvious singular behavior (e.g., mode conversion) when $\sin\tpb$ passes through zero in the $\varphi=0$ case (around $a\approx 6$, as seen by the nearly flat $b_{\|}$ at this time).
 
 Overall, we see tentative evidence that $\ph^{\rm (Y)}$ waves have a tendency to  generate larger compressive variations
 than $\ph^{\rm (Z)}$ waves. This could enable 
 other dissipation mechanisms in a real plasma thereby reducing the switchbacks generated by such waves. 
  Thus, this strengthens our conclusions from \cref{sec: SB formation due to expansion}, implying that, 
 as well as naturally generating smaller $b_{\|}$ due to geometry, switchbacks that involve field rotations in the normal direction
 also are likely to dissipate more strongly, thus enhancing the dominance of tangential switchbacks.

\section{The scaling of WKB waves in a strong Parker spiral}\label{app: zpx sclaing}

In deriving the scaling of wave amplitudes with expansion in the presence of a  Parker 
spiral $v_{{\rm A}y}\sim v_{{\rm A}x}$, we used the same scaling of the wave amplitude 
with expansion, $\overline{|\ba|}\propto a^{-1/2}$, as in the radial-background-field case. 
In this appendix, we confirm that this is indeed correct, \emph{viz.,} that the scaling of the unnormalized amplitude  of WKB Alfv\'en waves 
with $a$ is independent of their direction of propagation. This property has been shown in a number
of previous works for small amplitude waves \cite{Whang1973,Voelk1973a} and is also contained in the large-amplitude
results of Refs.~\onlinecite{Barnes1974,Hollweg1974} and \alfreds\ (the second term $(\dot{a}/2a)\ba$ in \cref{eq: alfreds b eqn} is independent of $\vah$).
Nonetheless, we feel that the EBM derivation 
below is helpful both for its generality (it does not assume 1-D waves or low amplitude) and its simplicity, which 
helps to illustrate the physical cause of the $\vah$-independence of the amplitude scaling.

Starting from the EBM equations (see e.g., \zades), we assume constant $\rho$ and incompressible motions, 
then form the equations for $\bm{z}^{\pm}=\bm{u}\pm\bm{B}/\sqrt{4\pi\rho}= \bm{u} \pm \ba\pm \va$
(where, as above, $\ba$ is $\delta \B/\sqrt{4\pi \rho}$). 
This gives 
\begin{align}
\partial_{t}\bm{z}^{\pm}\pm \va\cdot\tilde{\nabla}\bm{z}^{\pm} =& - \tilde{\nabla}\tilde{p} + \bm{z}^{\mp}\cdot\tilde{\nabla}\bm{z}^{\pm} \nonumber \\ &-\frac{\dot{a}}{2a}(z^{\pm}_{x}-z^{\mp}_{x})\hat{\bm{x}} -  \frac{\dot{a}}{2a} \bm{\mathsf{T}}\cdot (\bm{z}^{+}+\bm{z}^{-}),\label{eq: zpm equation x dep}
\end{align}
where $\bm{\mathsf{T}} = {\rm diag}(0,1,1)$, $\tilde{\nabla}$ is the $\nabla$ operator in the expanding frame, and $\tilde{p}$ is chosen to constrain  $\tilde{\nabla}\cdot\bm{z}^{\pm}=0$. The final two terms arise due to the differing influence of  expansion 
on $\bm{b}$ (the second-to-last term) and $\bm{u}$ (the last term).
To understand the scaling in the WKB limit neglecting nonlinear interactions, we can simply set $\bm{z}^{-}$ to zero. This
is justified because when the $\va\cdot\tilde{\nabla}$ term dominates the others, the reflection terms (those involving $\dot{a}/a\,\bm{z}^{+}$), cannot cause $\bm{z}^{-}$ grow because it quickly moves out of phase with the source $\bm{z}^{+}$ wave. Thus, in this limit, $(\partial_{t} + \va\cdot\tilde{\nabla})\bm{z}^{+}\approx - (\dot{a}/2a) \bm{z}^{+}$ because both the third and forth terms on the right-hand-side of \cref{eq: zpm equation x dep} involve the same factor $-\dot{a}/2a$. This implies that $|\bm{z}^{+}|$ decays as $|\bm{z}^{+}|\propto a^{-1/2}$, no matter its direction or the direction of the mean field.

\bibliographystyle{apsrev}
\bibliography{fullbib_formatted}


\end{document}